\documentclass[conference,9pt]{IEEEtran}
\IEEEoverridecommandlockouts

\usepackage{graphicx} 
\usepackage{braket}
\usepackage{subfigure}

\usepackage{xcolor}

\usepackage{hyperref}

\usepackage{tikz}
\definecolor{custompurple}{rgb}{0.5, 0.1, 0.5}
\usepackage{circuitikz}
\usetikzlibrary{quantikz}
\usepackage{setspace} 
\usepackage{circledsteps}

\pgfkeys{/csteps/inner ysep=10pt}
\usepackage{yquant}
\useyquantlanguage{groups}

\usepackage[normalem]{ulem}

\usepackage{amsmath}
\usepackage{amssymb}
\usepackage{amsfonts}

\usepackage{algorithm}
\usepackage{algorithmicx}
\usepackage{algpseudocode}

\newtheorem{defn}{\textbf{Definition}}
\newtheorem{exam}{Example}

\newtheorem{observ}{\textbf{Basic Construction}}

\newcommand {\F} {{\mathcal{F}}}

\newcommand {\low} {{\texttt{low}}}
\newcommand {\high} {{\texttt{high}}}
\newcommand {\idx} {{\texttt{idx}}}
\newcommand {\w} {{\texttt{wt}}}

\usepackage{authblk}

\def\BibTeX{{\rm B\kern-.05em{\sc i\kern-.025em b}\kern-.08em
    T\kern-.1667em\lower.7ex\hbox{E}\kern-.125emX}}

\begin{document}

\title{Advancing Quantum State Preparation Using Decision Diagram with Local Invertible Maps
\thanks{Work partially supported by Innovation Program for Quantum Science and Technology under Grant No. 2024ZD0300502,  Beijing Nova Program Grant No. 20220484128 and 20240484652, and the National Natural Science Foundation of China Grant No. 12471437.}}

\author[1]{Xin Hong}
\author[2]{Aochu Dai}
\author[1]{Chenjian Li}
\author[3]{Sanjiang Li}
\author[1]{Shenggang Ying}
\author[3]{Mingsheng Ying}
\affil[1]{Key Laboratory of System Software (Chinese Academy of Sciences) and State Key Laboratory of Computer Science}
\affil[ ]{Institute of Software, Chinese Academy of Sciences, Beijing, China}
\affil[2]{Department of Computer Science and Technology, Tsinghua University, Beijing, China}
\affil[3]{Centre for Quantum Software and Information, University of Technology Sydney, Sydney, Australia}
\affil[ ]{Emails: \{hongxin, licj, yingsg\}@ios.ac.cn, dac22@mails.tsinghua.edu.cn, \{sanjiang.li, mingsheng.ying\}@uts.edu.au}
\renewcommand\Authands{ and }

\maketitle

\begin{abstract}
Quantum state preparation (QSP) is a fundamental task in quantum computing and quantum information processing. It is critical to the execution of many quantum algorithms, including those in quantum machine learning. In this paper, we propose a family of efficient QSP algorithms tailored to different numbers of available ancilla qubits — ranging from no ancilla qubits, to a single ancilla qubit, to a sufficiently large number of ancilla qubits. Our approach exploits the power of Local Invertible Map Tensor Decision Diagrams (LimTDDs) — a highly compact representation of quantum states that combines tensor networks and decision diagrams to reduce quantum circuit complexity. Extensive experiments demonstrate that our methods significantly outperform existing approaches and exhibit better scalability for large-scale quantum states, both in terms of runtime and gate complexity. Furthermore, our method shows exponential improvement in best-case scenarios.  

\end{abstract}


\section{Introduction}

Quantum computing represents a transformative leap in computation capabilities, by leveraging the principles of quantum mechanics to achieve exponential speedups in specific computational tasks \cite{shor1994,mcardle2020quantum}.
One of the fundamental challenges in quantum computing and quantum information processing is the efficient preparation of quantum states, known as Quantum State Preparation (QSP). The success of many quantum algorithms, for example, quantum machine learning \cite{biamonte2017quantum} and HHL \cite{harrow2009quantum}, requires encoding classical data into quantum states efficiently. However, as the number of qubits in a quantum system increases, the complexity of representing and manipulating quantum states also grows exponentially, making efficient preparation of quantum states a highly challenging problem, particularly for large-scale quantum systems.

Significant progress has been made in recent years in developing efficient algorithms for QSP. Various methods have been proposed to prepare states of special types, such as sparse quantum states \cite{gleinig2021efficient,mao2024toward,PhysRevA.110.032609,luo2025space,li2024nearly}. Techniques based on gate decomposition \cite{plesch_quantum-state_2011,shende_synthesis_2006,Iten_Colbeck_Kukuljan_Home_Christandl_2016}, uniformly controlled rotations \cite{mottonen2004transformation}, and divide-and-conquer method \cite{araujo2021divide} have been developed for the preparation of general quantum states. Some works focus on minimising the number of ancilla qubits \cite{bergholm_quantum_2005}, while others aim to minimise the circuit depth \cite{zhang_quantum_2022,gui2024spacetime}, or find a trade-off between them \cite{zhang2021low}. Theoretical bounds for QSP have also been established \cite{sun2023asymptotically}. However, most existing methods rely on explicit representations of quantum states, such as vector representations, which grow exponentially as the number of qubits increases, limiting the scale of quantum states that can be efficiently prepared.

Decision diagrams have emerged as a powerful tool for efficiently representing and manipulating quantum states, primarily due to their compactness. Originally developed for classical computation, these data structures have been widely adopted in classical circuit synthesis \cite{5227151} and verification \cite{9707932}. Their application has since expanded into quantum computing, where researchers have adapted classical decision diagrams to support quantum operations or designed new variants. This has led to notable progress in the simulation and verification of quantum circuits \cite{burgholzer2021qcec,ftdd10528748}. 

More recently, decision diagrams have been applied to QSP \cite{mozafari_automatic_2020,mozafari_efficient_2022,tanaka_quantum_2024}. Leveraging their compactness, these algorithms enable the preparation of relatively large quantum states, mitigating the exponential memory cost associated with quantum state vectors. For instance, Mozafari et al. proposed an efficient algorithm for preparing uniform quantum states \cite{mozafari_automatic_2020}, while subsequent work further optimised the preparation process using one or more ancilla qubits \cite{mozafari_efficient_2022,tanaka_quantum_2024}. In such approaches, the compression efficiency of the underlying decision diagrams is a critical factor that  directly impacts the overall performance of the preparation algorithm.

A notable recent advancement in this area is the development of decision diagram enhanced with Local Invertible Maps (LIMs). Two such structures are the Local Invertible Map Decision Diagram (LIMDD) \cite{vinkhuijzen2023limdd}  and its extension to tensor computation---the Local Invertible Map Tensor Decision Diagram (LimTDD) \cite{hong_limtdd_2025}. LIMDD extends traditional decision diagrams by attaching LIMs to edges, where each LIM is a Pauli stabiser (i.e., a tensor product of Pauli operators possibly scaled by a global phase). This structure offers efficient representation of stabiliser states and Clifford circuits by identifying and exploiting isomorphisms between quantum states. 

LimTDD uses a richer class of local transformations known as XP-stabilisers, which include products of Pauli $X$ and diagonal $P$ operators with arbitrary phase factors. This generalisation allows LimTDD to represent a broader class of quantum states with even higher compression. Furthermore, LimTDD is designed to represent arbitrary tensors, enabling native support for tensor network structures. In fact, LimTDD can achieve exponential compression improvements over existing decision diagrams like TDD \cite{hong2020tensor} and LIMDD in the best cases. Additionally, the LimTDD software package (available at \url{https://github.com/Veriqc/LimTDD}) supports tensor operations like addition and contraction, making it a strong candidate for unifying quantum state preparation with circuit synthesis and verification workflows. While this paper focuses primarily on LimTDD, the QSP algorithms we propose are also fully compatible with LIMDD.

This paper presents efficient QSP algorithms based on LimTDD, which offers superior compression through its use of XP-stabilizers. We first propose an ancilla-free algorithm, then extend it to variants using one or multiple ancilla qubits, inspired by prior decision-diagram-based approaches \cite{mozafari_efficient_2022, tanaka_quantum_2024}. The time and gate complexities scale with the number of (reduced) paths and nodes in the LimTDD, respectively. Since LimTDDs typically contain fewer of both, our methods achieve substantial reductions in gate count. All algorithms operate via a top-down elimination of edge operators followed by a bottom-up weight adjustment. The first two follow a depth-first traversal; the third uses breadth-first. 

We provide an open-source tool that converts quantum state vectors into LimTDDs and synthesizes executable circuits. Our tool and algorithms are publicly available at \url{https://github.com/Veriqc/LimTDD_QSP}. Our experiments show that the resulting circuits are significantly more efficient for large-scale states. This work demonstrates the value of LimTDD in QSP and lays the groundwork for integration into quantum software platforms such as Qiskit, streamlining the workflow for both theorists and experimentalists.

This paper is a significant extension of \cite{xin2025limtddqsp}, where the algorithm uses one ancilla qubit was presented. This paper also proposes three algorithms with no or more ancilla qubits.

The structure of this paper is as follows. In section~\ref{sec:background}, we provide basic concepts of quantum computing, QSP, and LimTDD. In section~\ref{sec:basic_cons}, we introduce the basic constructions for QSP using LimTDD. From section~\ref{sec:Algorithms_mul} to \ref{sec:Algorithms_noa}, we give algorithms for QSP using LimTDD with no ancilla qubits, one ancilla qubit, sufficient number of ancilla qubits, and a given number of ancilla qubits. Then, we conduct experiments to carefully analyse the performance of our algorithm in section~\ref{sec:experiments}. Finally, in section~\ref{sec:conclusion}, we give a brief conclusion on our paper.

\section{Background}\label{sec:background}

In this section, we give basic background on quantum computing, quantum state preparation, and LimTDD.

\subsection{Quantum Computing}

\subsubsection{Quantum States}
Quantum computing harnesses the principles of quantum mechanics to perform computations using quantum bits (qubits), which can exist in superpositions of states, unlike classical bits that are either $0$ or $1$. A qubit is described by a two-dimensional complex vector space with orthonormal basis states $\ket{0}$ and $\ket{1}$, satisfying:

\[
\langle 0 | 0 \rangle = \langle 1 | 1 \rangle = 1 \quad \text{and} \quad \langle 0 | 1 \rangle = \langle 1 | 0 \rangle = 0.
\]

A general qubit state can be in a linear combination $\ket{\psi} = \alpha \ket{0} + \beta \ket{1}$, where $\alpha$ and $\beta$ are complex numbers with $|\alpha|^2 + |\beta|^2 = 1$. This superposition allows qubits to represent multiple states simultaneously and parallelly, offering an advantage over classical bits.

For multi-qubit systems, the state space grows exponentially with the number of qubits. An $n$-qubit system is described by a $2^n$-dimensional complex vector, with basis states $\ket{k}$ (binary strings of length $n$). For example, $\ket{000}$ represents all three qubits in the $\ket{0}$ state
$\ket{000} = \ket{0} \otimes \ket{0} \otimes \ket{0}.$

\begin{exam}\label{exp:q_state}
Consider the 3-qubit state $\frac{1}{\sqrt{6}}(\ket{000}+\ket{001}+\frac{1}{\sqrt{2}}\ket{010}-\frac{1}{\sqrt{2}}\ket{011}-\ket{100}-\frac{1}{\sqrt{2}}\ket{101}+\frac{1}{\sqrt{2}}\ket{110}+\ket{111})$, represented as an 8-dimensional vector: $\frac{1}{\sqrt{6}}[1,1,\frac{1}{\sqrt{2}},-\frac{1}{\sqrt{2}},-1,-\frac{1}{\sqrt{2}},\frac{1}{\sqrt{2}},1]^T$.
\end{exam}

\subsubsection{Quantum Gates}
Quantum gates, which are unitary transformations on qubits, perform quantum computations. Common gates include:

\begin{itemize}
    \item \textbf{Hadamard gate ($H$ gate)}: Creates superposition states:
    \[
    H\ket{0} = \frac{1}{\sqrt{2}}(\ket{0} + \ket{1}), \quad H\ket{1} = \frac{1}{\sqrt{2}}(\ket{0} - \ket{1}).
    \]
    
    \item \textbf{Pauli-$Z$ gate ($Z$ gate)}: Introduces a phase flip to $\ket{1}$:
    \[
    Z\ket{0} =\ket{0}, \quad Z\ket{1}=-\ket{1}.
    \]

    \item \textbf{Controlled-$X$ gate ($CX$ gate)}: Flips the target qubit if the control qubit is $\ket{1}$:
    \[
    CX(\ket{0} \otimes \ket{\psi}) = \ket{0} \otimes \ket{\psi}, \quad CX(\ket{1} \otimes \ket{\psi}) = \ket{1} \otimes X\ket{\psi}.
    \]
\end{itemize}

\subsubsection{Quantum Circuits}

Quantum circuits implement quantum algorithms by sequencing these gates. Gates are applied to qubits, and their order determines the overall transformation. Outputs are typically measured in the computational basis. Circuits can be graphically represented, with qubits as lines and gates as symbols.

\begin{figure}[htbp]
\centering
\begin{tikzpicture}
  \begin{yquant}[register/minimum height=9mm, operator/separation=5mm, control style={radius=2pt}, subcircuit box style={dashed}]
    qubit {$\ket{q_2}$} q2;
    qubit {$\ket{q_1}$} q1;
    qubit {$\ket{q_0}$} q0;
    box {$H$} q2;
    box {$H$} q1;
    box {$H$} q0;
    z q0 | q1;
    z q0 | q2;
    z q1 | q2;
  \end{yquant}
\end{tikzpicture}
\caption{Example of a quantum circuit with Hadamard and $CZ$ gates.}
    \label{fig:quantum_circuit}
\end{figure}
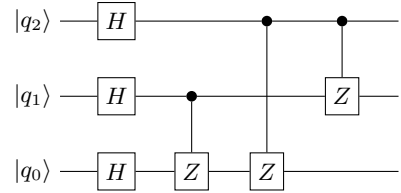

Fig. \ref{fig:quantum_circuit} shows a quantum circuit with Hadamard and $CZ$ gates, creating superposition and entanglement among three qubits.

\subsection{Quantum State Preparation}

The task of preparing a specific quantum state is a cornerstone in quantum computing and quantum information processing. This procedure is vital for executing various quantum algorithms, many of which demand particular quantum states as inputs to harness their computational benefits.

\subsubsection{Formal Definition}

Starting with a given initial state, commonly $\ket{0}^{\otimes n}$, the goal of quantum state preparation is to reach a target quantum state $\ket{\psi_v} = \sum_{k=0}^{2^n-1} v_k \ket{k}$. Here, $v = (v_0, v_1, \dots, v_{2^n-1})^T \in \mathbb{C}^{2^n}$ is a normalized vector ($\|v\|_2 = 1$) that encapsulates the amplitudes of the target state in the computational basis. The mission of QSP is to devise a quantum circuit capable of converting the initial state into the target state $\ket{\psi_v}$.

\subsubsection{Key Challenges and Significance}

In the general case, the complexity of representing and manipulating quantum states escalates exponentially with the number of qubits $n$. To fully characterise a general $n$-qubit state, $2^n$ complex amplitudes are necessary, rendering the explicit representation and preparation of arbitrary quantum states computationally prohibitive in most scenarios. This makes efficient QSP an exceedingly challenging problem, particularly for large-scale quantum systems. Nevertheless, it is indispensable for practical quantum computing applications, as the efficiency of QSP directly affects the viability and performance of numerous quantum algorithms.

\subsection{LimTDD}\label{subsec:LimTDD}


LimTDD is an advanced decision diagram designed for the efficient representation and manipulation of tensors and tensor networks, with its compression efficiency grounded in the concept of quantum state isomorphism.

\vspace{0.2em}
\begin{defn}[LIM, Quantum State Isomorphism \cite{vinkhuijzen2023limdd}]
An $n$-qubit Local Invertible Map (LIM) is an operator  
\begin{equation}\label{eq:limO}
    O = \lambda\, O_{n-1} \otimes \cdots \otimes O_0, 
\end{equation} 
where $\lambda\in \mathbb C$ is a complex number and each $O_i$ is an invertible $2\times 2$ matrix. The set of all such maps is denoted as $\mathcal{M}(n)$, and the set of all LIMs is defined as 
\begin{equation}\label{eq:limCalO}
\mathcal{M} = \bigcup_{n\in N}{\mathcal{M}(n)}.
\end{equation} 
Two $n$-qubit quantum states $\ket{\Psi}$ and $\ket{\Phi}$ are said to be \emph{isomorphic} if $\ket{\Phi} = O \ket{\Psi}$ for some $O \in \mathcal{M}(n)$.
\end{defn}
Special cases of LIMs include Pauli operators and XP-operators (cf.~\cite{vinkhuijzen2023limdd,hong_limtdd_2025}), with the latter being more general than the former.

\vspace{0.2em}
\begin{defn}[LimTDD \cite{vinkhuijzen2023limdd,hong_limtdd_2025}]\label{def:limtdd}
	Let $\mathcal{G}$ be a subgroup of $\mathcal{M}$. A $\mathcal{G}$-LimTDD $\mathcal{F}$ over a set of indices $S$ is a rooted, weighted, and directed acyclic graph 
	$\mathcal{F} = (V, E, \idx, \low, \high, \w)$ defined as follows:
	\begin{itemize}
		\item $V$ is a finite set of nodes which consists of non-terminal nodes $V_{NT}$ and a terminal node $v_T$ labelled with integer 1. Denote by $r_\mathcal{F}$ the unique root node of $\mathcal{F}$;
		\item $\idx: V_{NT} \rightarrow S$ assigns each non-terminal node an index in $S$. We call $\idx(r_\mathcal{F})$ the top index of $\F$, if $r_\mathcal{F}$ is not the terminal node;
		\item both $\low$ and $\high$ are mappings in $V_{NT} \rightarrow V$, which map each non-terminal node to its 0- and 1-successors, respectively;
		\item $E = \{(v, \low(v)), (v, \high(v)) : v\in V_{NT}\}$ is the set of edges, where $(v, \low(v))$ and $(v, \high(v)) $ are called the low- and high-edges of $v$, respectively. For simplicity, we also assume the root node $r_\mathcal{F}$ has a unique incoming edge, denoted  $e_r$, which has no source node;
		\item $\w: E\rightarrow \mathcal{G}$ assigns each edge a weight in $\mathcal{G}$.  $\w(e_r)$ is called the weight of $\mathcal{F}$, and denoted $w_\mathcal{F}$. 
	\end{itemize} 

\end{defn}

LimTDD is an extension of LIMDD \cite{vinkhuijzen2023limdd}. While the two have the same form, they differ in their implementations and semantics. LIMDD exploits Pauli operators to represent and process quantum states, but LimTDD uses more general XP-operators to represent and process tensors. In this paper, we are only concerned with its application in representing quantum states. In this case, the semantics of the terminal node is defined to be $\ket{v_T} = 1$, the semantics of an edge $e$, directing to a node $v$, is defined as
$$
\ket{e} = \w(e) \cdot \ket{v},
$$
and the semantics of a non-terminal node $v$ is defined to be 
$$
\ket{v} = \ket{0}\otimes \big|(v,\low(v))\big\rangle + \ket{1}\otimes \big|(v,\high(v))\big\rangle.\\
$$

In this paper, each index corresponds to a qubit, and we will use $q_v$ to represent the qubit corresponding to a node $v$. For convenience, we assume that the top index represents the most significant qubit ($q_{n-1}$ for an $n$-qubit state), and the bottom non-terminal node's index represents the least significant qubit ($q_0$), with other qubits arranged sequentially. Occasionally, we use notations like $\ket{0}_{n-1}$ and $\ket{0}_0$ to identify specific qubits, and we will use notation $\ket{0}_v$ to represent a $\ket{0}$ state on the qubit corresponding to the node $v$. In addition, we use will $\ket{*}_n$ to represent all possible computational bases of $n$ qubits, and we will omit $n$ if no ambiguity. The subgroup $\mathcal{G}$ is set to XP-Operators in \cite{hong_limtdd_2025}, but our algorithm applies to any subgroup of $\mathcal{M}$.


\begin{figure}[htbp]
    \centering
    \resizebox{0.2\textwidth}{!}{




\begin{tikzpicture}[
>=latex,
line join=bevel,
every node/.style={minimum size=1.4cm, inner sep=0pt,thick, font=\LARGE}, 
every path/.style={line width=1.5pt} 
]
\node (1) at (128.67bp,18.0bp) [draw=red,circle,font=\Huge] {$1$};
  \node (y01) at (46.668bp,108.9bp) [draw=red,circle,font=\Huge] {$v_{00}$};
  \node (y02) at (185.67bp,108.9bp) [draw=red,circle,font=\Huge] {$v_{01}$};
  \node (y10) at (66.668bp,202.96bp) [draw=red,circle,font=\Huge] {$v_{10}$};
  \node (y11) at (165.67bp,202.96bp) [draw=red,circle,font=\Huge] {$v_{11}$};
  \node (y2) at (107.67bp,297.02bp) [draw=red,circle,font=\Huge] {$v_{20}$};
  \node (-0) at (107.67bp,387.93bp) [draw,draw=none] {};
  \draw [red,->,dotted] (y01) ..controls (29.551bp,79.314bp) and (25.128bp,64.774bp)  .. (32.668bp,54.0bp) .. controls (47.583bp,32.686bp) and (76.911bp,24.306bp)  .. (1);
  \definecolor{strokecol}{rgb}{0.0,0.0,0.0};
  \pgfsetstrokecolor{strokecol}
  \draw (53.668bp,61.875bp) node {};
  \draw [blue,->] (y01) ..controls (73.917bp,78.36bp) and (94.182bp,56.39bp)  .. (1);
  \draw (116.67bp,61.875bp) node {};
  \draw [red,->,dotted] (y02) ..controls (166.06bp,77.316bp) and (153.94bp,58.425bp)  .. (1);
  \draw (180.67bp,61.875bp) node {};
  \draw [blue,->] (y02) ..controls (205.77bp,80.723bp) and (212.16bp,65.699bp)  .. (204.67bp,54.0bp) .. controls (194.36bp,37.899bp) and (174.56bp,29.102bp)  .. (1);
  \draw (230.29bp,61.875bp) node {$\frac{1}{\sqrt{2}}$};
  \draw [red,->,dotted] (y10) ..controls (31.234bp,190.19bp) and (11.932bp,180.12bp)  .. (2.6676bp,163.81bp) .. controls (-5.0407bp,150.24bp) and (6.2358bp,136.64bp)  .. (y01);
  \draw (23.668bp,155.93bp) node {};
  \draw [blue,->] (y10) ..controls (59.707bp,169.93bp) and (56.313bp,154.3bp)  .. (y02);
  \draw (105.543bp,155.93bp) node {$\frac{1}{\sqrt{2}} S$};
  \draw [red,->,dotted] (y11) ..controls (163.61bp,171.29bp) and (164.02bp,158.88bp)  .. (166.67bp,148.06bp) .. controls (167.5bp,144.65bp) and (168.68bp,141.19bp)  .. (y02);
  \draw (187.67bp,155.93bp) node {};
  \draw [blue,->] (y11) ..controls (193.01bp,184.7bp) and (203.51bp,175.28bp)  .. (208.67bp,163.81bp) .. controls (212.85bp,154.5bp) and (210.23bp,144.22bp)  .. (y02);
  \draw (230.42bp,155.93bp) node {$X$};
  \draw [red,->,dotted] (y2) ..controls (93.74bp,264.75bp) and (86.177bp,247.77bp)  .. (y10);
  \draw (110.67bp,249.99bp) node {};
  \draw [blue,->] (y2) ..controls (125.36bp,273.0bp) and (130.96bp,265.2bp)  .. (135.67bp,257.87bp) .. controls (140.87bp,249.76bp) and (146.13bp,240.69bp)  .. (y11);
  \draw (172.92bp,249.99bp) node {$Z\otimes I$};
  \draw [red,->] (-0) ..controls (107.67bp,358.85bp) and (107.67bp,343.46bp)  .. (y2);
  \draw (154.54bp,344.05bp) node {$\frac{2}{\sqrt{23}} Z\otimes I\otimes I$};
\end{tikzpicture}

%
%
}
    \caption{An example of LimTDD representing the quantum state $\ket{\F} = \frac{2}{\sqrt{23}}[1,1,\frac{1}{\sqrt{2}},\frac{i}{2},-1,-\frac{1}{\sqrt{2}},\frac{1}{\sqrt{2}},1]^T$. We omit the weight $1$ and $1\cdot I^{\otimes k}$ in the figure. 
    }
    \label{fig:limtdd}
\end{figure}
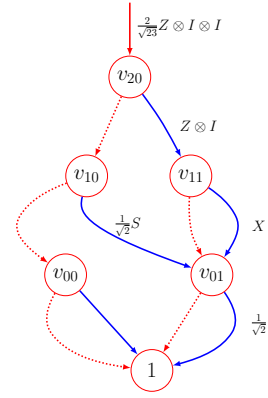

\begin{exam}\label{exp:limtdd}
Fig. \ref{fig:limtdd} illustrates an example of a LimTDD representing the quantum state from Example \ref{exp:q_state}. In this diagram, low edges are depicted with dotted red lines and high edges with solid blue lines. The qubit corresponds to node $v_{20}$ is $q_2$, the qubit corresponds to node $v_{10}$ and $v_{11}$ is $q_1$, and the qubit corresponds to node $v_{00}$ and $v_{01}$ is $q_0$. Ignoring normalisation coefficients:
\begin{itemize}
    \item The $v_{00}$ node represents $\ket{v_{00}} = \ket{0} + \ket{1}$, and the $v_{01}$ node represents $\ket{v_{01}} = \ket{0} + \frac{1}{\sqrt{2}} \ket{1}$.
    \item The $v_{10}$ node represents $\ket{v_{10}} = \ket{0}\ket{v_{00}} + \frac{1}{\sqrt{2}}\ket{1}(S \ket{v_{01}}) = \ket{00} + \ket{01} + \frac{1}{\sqrt{2}}\ket{10} + \frac{i}{2}\ket{11}$, and the $v_{11}$ node represents $\ket{v_{11}} = \ket{0}\ket{v_{01}} + \ket{1}(X \ket{v_{01}}) = \ket{00} + \frac{1}{\sqrt{2}}\ket{01} + \frac{1}{\sqrt{2}}\ket{10} + \ket{11}$.
    \item The $v_{20}$ node represents $\ket{v_{20}} = \ket{0}\ket{v_{10}} + \ket{1}(Z \otimes I \ket{v_{11}}) = \ket{000} + \ket{001} + \frac{1}{\sqrt{2}}\ket{010} + \frac{i}{2}\ket{011} + \ket{100} + \frac{1}{\sqrt{2}}\ket{101} - \frac{1}{\sqrt{2}}\ket{110} - \ket{111}$.
    \item The entire LimTDD represents the quantum state $\frac{2}{\sqrt{23}} Z \otimes I \otimes I \ket{v_{20}} = \frac{2}{\sqrt{23}}(\ket{000} + \ket{001} + \frac{1}{\sqrt{2}}\ket{010} + \frac{i}{2}\ket{011} - \ket{100} - \frac{1}{\sqrt{2}}\ket{101} + \frac{1}{\sqrt{2}}\ket{110} + \ket{111})$. For convience, we denote the quantum state as $\ket{\F}$.
\end{itemize}
\end{exam}

\vspace{0.2em}

We will also use the following graphical notation to describe LimTDDs. Each node within a LimTDD is uniquely characterised by three key elements: its index, the two nodes it points to (successors), and the weights assigned to the edges leading to these successors. This relationship can be visually represented as:
\[
\Circled{v_0} \overset{w_0}{\dashleftarrow} \Circled{v} \xrightarrow{w_1} \Circled{v_1}.
\]

Moreover, the entire LimTDD structure is uniquely defined by its root node and the weight associated with the edge entering this root node. This can be expressed as:
\[
\left(w_\mathcal{F},\ \Circled{v_0} \overset{w_0}{\dashleftarrow} \Circled{v} \xrightarrow{w_1} \Circled{v_1}\right),
\]
or more compactly as:
\[
\xrightarrow{w_\mathcal{F}} \Circled{v}.
\]

Note that, for a LimTDD representing a quantum state, when normalisation has been applied, every non-terminal node will have the following form:
\[
\Circled{v_0} \overset{I}{\dashleftarrow} \Circled{v} \xrightarrow{\lambda\cdot O} \Circled{v_1}.
\]

For any unnormalised quantum state \( \ket{\psi} \), there exists a unitary operator \( U \) such that \( U\ket{\psi} = w\ket{0} \), where \( w = \sqrt{\braket{\psi|\psi}} \) represents the 2-norm of \( \ket{\psi} \). When a node \( \Circled{v_0} \overset{\lambda_0 \cdot O}{\dashleftarrow} \Circled{v} \xrightarrow{\lambda_1 \cdot O'} \Circled{v_1} \) represents the quantum state \( \ket{\psi} \), where \( \lambda_0 \) and \( \lambda_1 \) are two complex numbers, the 2-norm of \( v \), denoted as \( ||v|| \), is defined to be \( \sqrt{\braket{\psi|\psi}} \), which can be computed based on the norms of its two successors. Specifically,
\[
||v|| = \sqrt{|\lambda_0|^2 \cdot ||v_0||^2 + |\lambda_1|^2 \cdot ||v_1||^2},
\]
with the norm of the terminal node \( ||v_T|| \) set to 1.

In this paper, we use bit-strings \( b_{n-1}\cdots b_0 \) (where each \( b_i \in \{0,1\} \)) to represent paths in a LimTDD. The prefix path leading to a node \( v \) along a path \( b_{n-1}\cdots b_0 \) is defined as the prefix \( b_{n-1}\cdots b_{k+1} \), if the qubit corresponding to \( v \) is \( q_k \). The product of all the weights along a (prefix) path is called an accumulated weight.

\begin{exam}\label{exp:limtdd}
In Fig. \ref{fig:limtdd}, the left-most path (red-red-red) is 000. The prefix path leading to the node \( v_{10} \) along this path is 0, and the prefix path leading to the node \( v_{00} \) is 00, where the accumulated weights along both paths are \( \frac{1}{\sqrt{23}}Z\otimes I\otimes I \).
\end{exam}

\section{Basic Constructions} \label{sec:basic_cons}

The methods for QSP using LimTDD are mainly based on the following basic construction components.

\subsection{Basic Construction Components}


\begin{observ}[Incoming Edge Operator Elimination]
    Let $\mathcal{F} = \xrightarrow{\lambda O_n \otimes \cdots \otimes O_1} \Circled{v}$ be a LimTDD representing the quantum state $\ket{\psi}$. Then, we have
    \[
    \ket{\psi} = \lambda (O_n \otimes \cdots \otimes O_1) \ket{v}.
    \]
    Applying the operator $(O_n \otimes \cdots \otimes O_1)^\dag$ to $\ket{\psi}$ reduces it to $\lambda \ket{v}$, which can be represented by a LimTDD with the root node $v$ and incoming edge weight $\lambda$, that is,
    \[
    \xrightarrow{\lambda} \Circled{v}.
    \]
    In other words, applying $(O_n \otimes \cdots \otimes O_1)^\dag$ to the LimTDD (i.e., contracting each index $x_i$ (corresponding to $q_i$) with an operator $O_i$) eliminates the operator on the incoming edge of the LimTDD.
\end{observ}\vspace{0.2em}




\begin{observ}[High-Edge operator Elimination]
Let
\[
\Circled{v_0} \overset{I}{\dashleftarrow} \Circled{v} \xrightarrow{\lambda O_n \otimes \cdots \otimes O_1} \Circled{v_1}
\]
be a non-terminal node of a LimTDD \(\mathcal{F}\). Let \(p\) be the prefix path leading to \(v\), and assume the accumulated weight along \(p\) is a complex number \(w\). Then, the quantum state represented by \(\mathcal{F}\) can be expressed as
$\ket{\psi} = w \cdot \ket{p}\ket{v} + \ket{\rm Res} = w \cdot \ket{p}\ket{0}\ket{v_0} + w \cdot \lambda \cdot \ket{p}\ket{1} (O_n \otimes \cdots \otimes O_1 \ket{v_1}) + \ket{\rm Res}$,
where \(\ket{\rm Res}\) is orthogonal to \(\ket{p}\ket{*}\). By applying the controlled operator \((O_n \otimes \cdots \otimes O_1)^\dag\) with the control condition \(\ket{p}\ket{1}\), the state is transformed into
\[
w \cdot \ket{p}\ket{0}\ket{v_0} + w \cdot \lambda \cdot \ket{p}\ket{1} \ket{v_1} + \ket{\rm Res}.
\]
Consequently, the node is updated to
$\Circled{v_0} \overset{I}{\dashleftarrow} \Circled{v'} \xrightarrow{\lambda} \Circled{v_1}$,
indicating that the operator on the high-edge of the node has been successfully eliminated.
\end{observ}
\vspace{0.2em}


\begin{observ}[Outgoing Weight Reduction]
Consider a non-terminal node of a LimTDD \(\mathcal{F}\):
\[
\Circled{v_0} \overset{w_0}{\dashleftarrow} \Circled{v} \xrightarrow{w_1} \Circled{v_0},
\]
where \(w_0\) and \(w_1\) are complex numbers with \(w_0 \neq 0\). Let \(p\) denote the prefix path leading to \(v\), and assume the accumulated weight along \(p\) is a complex number \(w\). The quantum state represented by \(\mathcal{F}\) can then be written as
$\ket{\psi} = w \cdot \ket{p}\ket{v} + \ket{\rm Res} = w \cdot \ket{p}(w_0\ket{0} + w_1\ket{1})\ket{v_0} + \ket{\rm Res}$
,where \(\ket{\rm Res}\) is orthogonal to \(\ket{p}\ket{*}\). Define \(c = w_1 / w_0\). Applying the controlled unitary operator
\[
\frac{1}{\sqrt{1 + |c|^2}}
\begin{bmatrix}
1 & c^\dag \\
-c & 1 \\
\end{bmatrix}
\]
to \(\ket{\psi}\) with the control condition \(\ket{p}\) transforms the state into
$w \cdot \sqrt{|w_0|^2 + |w_1|^2} \cdot \ket{0}\ket{v_0} + \ket{\rm Res}.$ 
Consequently, the node is updated to
$\Circled{v_0} \overset{1}{\dashleftarrow} \Circled{v'} \xrightarrow{0} \Circled{v_0}.$
This indicates that the complex weights on the two outgoing edges of \(v\) have been successfully reduced to \([1, 0]\), corresponding to \(\ket{0}\).
\end{observ}

It is important to note that the accumulated weight leading to the node must be a complex number, in both Constructions 2 and 3. Additionally, in Construction 3, both successor nodes must be identical. This implies that whenever we process a node, we must first eliminate all operators on the paths leading to it. When dealing with the outgoing complex weights of a node, we must first ensure that its two successors are transformed into the same state. Therefore, in the four proposed algorithms, we will first traverse the LimTDD from top to bottom to sequentially eliminate the operators. After reaching the bottom, we will then address the complex weights in reverse order. By the time we handle the outgoing weights of a node, both of its successors will have been converted to represent \(\ket{0}^{\otimes k}\) for some \(k\).

\subsection{Branch Condition}

Note that in Basic Constructions 2 and 3, we utilise the values of all qubits along a prefix path leading to the node \( v \) as the control condition. Typically, this can be simplified to using the values of all branch nodes (branch condition).

For instance, consider a quantum state 
\[
\ket{\psi} = \ket{0}_3 \ket{+}_2 \ket{1}_1 \ket{v} + \ket{\rm Res},
\]
where \(\ket{\rm Res}\) is some state orthogonal to \(\ket{0}_3 \ket{0}_2 \ket{1}_1 \ket{*}\) and \(\ket{0}_3 \ket{1}_2 \ket{1}_1 \ket{*}\). In this case, the operator on the high-edge of \( v \) can be eliminated using the control condition \(\ket{0}_3 \ket{1}_1 \ket{1}_v\), and the weights on the two outgoing edges of \( v \) can be reduced using the control condition \(\ket{0}_3 \ket{1}_1\).

\begin{defn}[Branch Condition]
A non-terminal node is termed a branch node if its 0-successor and 1-successor are distinct. The branch condition for a node along a path is defined by the values of all branch nodes preceding it on that path.
\end{defn}

Consider the LimTDD given in Fig.~\ref{fig:limtdd}. The nodes \( v_{20} \) and \( v_{10} \) are branch nodes. The branch condition for the \( v_{10} \) node along the path \( 000 \) is \(\ket{0}_2\), while for the \( v_{00} \) node, it is \(\ket{0}_2 \ket{0}_1\).

Through this operation, we effectively handle multiple paths simultaneously, which we collectively refer to as a reduced path.

\begin{defn}[Reduced paths]\label{def:red_path}
Let $\mathcal{F}$ be a decision diagram. The \emph{reduced diagram} of $\mathcal{F}$ is obtained by merging the edges between any two nodes in $\mathcal{F}$. We also call paths within this reduced diagram \emph{reduced paths} of $\mathcal{F}$.
\end{defn}

In the provided example, the node \( v_{11} \) has identical 0-successor and 1-successor nodes, meaning its two outgoing edges are merged when determining the reduced paths. The same applies to the nodes \( v_{00} \) and \( v_{01} \). Consequently, the LimTDD features only 3 reduced paths. The number of reduced paths is a primary indicator of complexity for some of the algorithms proposed below.

\subsection{Open/Closed Node}

It is important to note that while the branch condition is introduced to reduce the number of control qubits, it may still involve a significant number of qubits, thereby increasing the complexity of the resulting circuit. To mitigate this, ancilla qubits can be introduced to further reduce the number of control qubits.

For instance, consider the quantum state
$\ket{\psi} = w \cdot \ket{p}\ket{v} + \ket{\rm Res} = w \cdot \ket{p}\ket{0}\ket{v_0} + w \cdot \lambda \cdot \ket{p}\ket{1} (O_n \otimes \cdots \otimes O_1 \ket{v_1}) + \ket{\rm Res}$.
By introducing an ancilla qubit \( q_a \), we can adjust the quantum state to
$
w \cdot \ket{1}_a \ket{p} \ket{v} + \ket{0}_a \ket{\rm Res} = w \cdot \ket{1}_a \ket{p} \ket{0}_v \ket{v_0} + w \cdot \lambda \cdot \ket{1}_a \ket{p} \ket{1}_v (O_n \otimes \cdots \otimes O_1 \ket{v_1}) + \ket{0}_a \ket{\rm Res}.
$
Then, the operator \( O_n \otimes \cdots \otimes O_1 \) can be eliminated using the control condition \( \ket{1}_a \ket{1}_v \). In this case, we say that the node \( v \) is marked open by the ancilla qubit \( q_a \). More specifically:

\begin{defn}[Open/Closed Node]
Let \(\mathcal{F}\) be a LimTDD representing the quantum state \(\ket{\psi} = \sum_{k=0}^{2^n-1} \lambda_k \ket{k}\). By introducing an ancilla qubit \( q_a \) and marking each computational basis state \(\ket{k}\) with \(\ket{b_k}_a\), we obtain the state \(\sum_{k=0}^{2^n-1} \lambda_k \ket{b_k}_a \ket{k}\), where \( b_k \in \{0,1\} \). A path \( p \), corresponding to the state \(\ket{p}\), is called open (closed, respectively) by \( q_a \) if \( b_p = 1 \) (\(0\), respectively). A node \( v \) with prefix path \( p \) is called open (closed, respectively) by \( q_a \) if all states \(\ket{p}\ket{*}\) are marked open (closed). A LimTDD is called open (closed) if its root node is open (closed).
\end{defn}

In the following sections, we will introduce various strategies to control the opening and closing of nodes so that they can be processed without affecting other parts of the state (decision diagram).

\section{LimTDD based QSP with No Ancilla Qubit} \label{sec:Algorithms_noa}

Having explored the fundamental concepts, we now turn to the algorithms for efficient quantum state preparation (QSP).

\subsection{Algorithm}

We first consider the scenario where no ancilla qubit is available. In this case, all high-edge operators and outgoing weights must be eliminated or reduced using the branch condition. The procedure is detailed in Alg.~\ref{alg:State_Pre_noa}.

Before initiating the procedure, we apply Basic Construction 1 to eliminate the operator on the incoming edge of the LimTDD \(\mathcal{F}\), then execute the algorithm with the input \(r_{\mathcal{F}}\).

In this procedure, for each node \(v\) represented as
\[
\Circled{v_0} \overset{I}{\dashleftarrow} \Circled{v} \xrightarrow{\lambda O} \Circled{v_1},
\]
where \(\ket{v} = \ket{0}\ket{v_0} + \lambda \ket{1}(O\ket{v_1})\), we first address the operator \(O\) on the high-edge of \(v\) using Basic Construction 2. The state is transformed to
\[
\ket{v} = \ket{0}\ket{v_0} + \lambda \ket{1}\ket{v_1}.
\]
We then recursively process its successors \(v_0\) and \(v_1\). If \(v\) is a branch node, the circuits \(cir_0\) and \(cir_1\) obtained from processing \(v_0\) and \(v_1\) are appended to the main circuit with control conditions \(\ket{0}_v\) and \(\ket{1}_v\), respectively. Otherwise, if \(v_0 = v_1\) and \(cir_0 = cir_1\), we simply add this circuit to the preparation circuit without any additional control conditions. Suppose \(cir_0\) and \(cir_1\) transform \(\ket{v_0}\) and \(\ket{v_1}\) to \(||v_0||\ket{0}^{\otimes k}\) and \(||v_1||\ket{0}^{\otimes k}\) for some \(k\). Appending these circuits transforms the original state to
$(||v_0|| \ket{0} + \lambda ||v_1|| \ket{1}) \ket{0}^{\otimes k}$.
Subsequently, applying the operator
\[
\frac{1}{\sqrt{1 + |c|^2}}
\begin{bmatrix}
1 & c^\dag \\
-c & 1 \\
\end{bmatrix}
\]
with \(c = \lambda \cdot ||v_1|| / ||v_0||\) reduces the weight \([||v_0||, \lambda \cdot ||v_1||]\) to
$\sqrt{||v_0||^2 + |\lambda|^2 \cdot ||v_1||^2} \cdot [1, 0]$,
where \(\sqrt{||v_0||^2 + |\lambda|^2 \cdot ||v_1||^2} = ||v||\). Consequently, the node \(v\) is transformed to \(||v|| \ket{0}^{\otimes k+1}\) by the resulting circuit.

\begin{algorithm} 
\caption{$\textsc{StatePre1}(v)$}
\begin{algorithmic}[1]
\Require{A node $v$ of an LimTDD representing an $n$-qubit quantum state $\ket{\psi}$, suppose the qubit corresponding to $v$ is $q_v$. }
\Ensure{A quantum circuit $C$, corresponding to an unitary matrix $U$, such that $U \ket{\psi}= \ket{0}^{\otimes n}$.}
\vspace{0.4em}

\State $cir \gets \mathrm{QuantumCircuit}(n)$ {\color{gray}\Comment{An empty quantum circuit with $n$ qubits}}

\If{$v$ is the terminal node}
\State $||v|| \gets 1$
\State \Return $cir$
\EndIf
\State Suppose $\w\big((v,\high(v))\big)=\lambda \cdot O$
\State Append $cir$ with a controlled $O^\dag$ gate, with the control condition set to be $\ket{1}_v$ {\color{gray}\Comment{Reduce the operator on the high-edge of $v$}}

\If{$\low(v)=\high(v)$}
\State $cir_0 \gets \textsc{StatePre1}\big(\low(v)\big)$
\State Append $cir$ with $cir_0$
\Else
\State $cir_0 \gets \textsc{StatePre1}\big(\low(v)\big)$
\State Add a control qubit with control condition $\ket{0}_v$ for every gate in $cir_0$ and append the circuit to $cir$
\State $cir_1 \gets \textsc{StatePre1}\big(\high(v)\big)$
\State Add a control qubit with control condition $\ket{1}_v$ for every gate in $cir_1$ and append the circuit to $cir$

\EndIf
\State $w_0 \gets ||\low(v)||$
\State $w_1 \gets \lambda \cdot ||\high(v)||$
\State $c \gets w_1/w_0$
\State Append a gate $\frac{1}{\sqrt{1+|c|^2}}\left[\begin{array}{cccc} 
		1 & c^\dag\\ 
		-c & 1\\
\end{array}\right]$ to qubit $q_v$ in $cir$
\vspace{0.2em}
\State $||v|| \gets \sqrt{|w_0|^2+|w_1|^2}$
\State \Return $cir$
\end{algorithmic}
\label{alg:State_Pre_noa} 
\end{algorithm}

\subsection{An Example}\label{sec:exp}

\begin{figure*}[htbp]
\centering

\subfigure[]{
    \resizebox{0.2\textwidth}{!}{




\begin{tikzpicture}[
>=latex,
line join=bevel,
every node/.style={minimum size=1.4cm, inner sep=0pt,thick, font=\LARGE}, 
every path/.style={line width=1.5pt} 
]
\node (1) at (128.67bp,18.0bp) [draw=red,circle,font=\Huge] {$1$};
  \node (y01) at (46.668bp,108.9bp) [draw=red,circle,font=\Huge] {$v_{00}$};
  \node (y02) at (185.67bp,108.9bp) [draw=red,circle,font=\Huge] {$v_{01}$};
  \node (y10) at (66.668bp,202.96bp) [draw=red,circle,font=\Huge] {$v_{10}$};
  \node (y11) at (165.67bp,202.96bp) [draw=red,circle,font=\Huge] {$v_{11}$};
  \node (y2) at (107.67bp,297.02bp) [draw=red,circle,font=\Huge] {$v_{20}$};
  \node (-0) at (107.67bp,387.93bp) [draw,draw=none] {};
  \draw [red,->,dotted] (y01) ..controls (29.551bp,79.314bp) and (25.128bp,64.774bp)  .. (32.668bp,54.0bp) .. controls (47.583bp,32.686bp) and (76.911bp,24.306bp)  .. (1);
  \definecolor{strokecol}{rgb}{0.0,0.0,0.0};
  \pgfsetstrokecolor{strokecol}
  \draw (53.668bp,61.875bp) node {};
  \draw [blue,->] (y01) ..controls (73.917bp,78.36bp) and (94.182bp,56.39bp)  .. (1);
  \draw (116.67bp,61.875bp) node {};
  \draw [red,->,dotted] (y02) ..controls (166.06bp,77.316bp) and (153.94bp,58.425bp)  .. (1);
  \draw (180.67bp,61.875bp) node {};
  \draw [blue,->] (y02) ..controls (205.77bp,80.723bp) and (212.16bp,65.699bp)  .. (204.67bp,54.0bp) .. controls (194.36bp,37.899bp) and (174.56bp,29.102bp)  .. (1);
  \draw (230.29bp,61.875bp) node {$\frac{1}{\sqrt{2}}$};
  \draw [red,->,dotted] (y10) ..controls (31.234bp,190.19bp) and (11.932bp,180.12bp)  .. (2.6676bp,163.81bp) .. controls (-5.0407bp,150.24bp) and (6.2358bp,136.64bp)  .. (y01);
  \draw (23.668bp,155.93bp) node {};
  \draw [blue,->] (y10) ..controls (59.707bp,169.93bp) and (56.313bp,154.3bp)  .. (y02);
  \draw (105.543bp,155.93bp) node {$\frac{1}{\sqrt{2}} S$};
  \draw [red,->,dotted] (y11) ..controls (163.61bp,171.29bp) and (164.02bp,158.88bp)  .. (166.67bp,148.06bp) .. controls (167.5bp,144.65bp) and (168.68bp,141.19bp)  .. (y02);
  \draw (187.67bp,155.93bp) node {};
  \draw [blue,->] (y11) ..controls (193.01bp,184.7bp) and (203.51bp,175.28bp)  .. (208.67bp,163.81bp) .. controls (212.85bp,154.5bp) and (210.23bp,144.22bp)  .. (y02);
  \draw (230.42bp,155.93bp) node {$X$};
  \draw [red,->,dotted] (y2) ..controls (93.74bp,264.75bp) and (86.177bp,247.77bp)  .. (y10);
  \draw (110.67bp,249.99bp) node {};
  \draw [blue,->] (y2) ..controls (125.36bp,273.0bp) and (130.96bp,265.2bp)  .. (135.67bp,257.87bp) .. controls (140.87bp,249.76bp) and (146.13bp,240.69bp)  .. (y11);
  \draw (172.92bp,249.99bp) node {$Z\otimes I$};
  \draw [red,->] (-0) ..controls (107.67bp,358.85bp) and (107.67bp,343.46bp)  .. (y2);
  \draw (154.54bp,344.05bp) node {$\frac{2}{\sqrt{23}} Z\otimes I\otimes I$};
\end{tikzpicture}

%
%
}
}
\subfigure[]{
\begin{tikzpicture}
  \begin{yquant}[register/minimum height=17mm, operator/separation=3mm, control style={radius=2pt}, subcircuit box style={dashed}]
    qubit {$\ket{q_2}$} q2;
    qubit {$\ket{q_1}$} q1;
    qubit {$\ket{q_0}$} q0;
    box {$Z$} q2;
    z q1 | q2;
    subcircuit {
      qubit {} q2;
      qubit {} q1;
      qubit {} q0;
      box {$S^\dag$} q0 | q1, ~q2;
      box {$U$} q0 | ~q1, q2;
      box {$V$} q0 | q1, ~q2;
      box {$W$} q1 | ~q2;
    } (q2,q1,q0);
    subcircuit {
      qubit {} q2;
      qubit {} q1;
      qubit {} q0;
      x q0 | q1,q2;
      box {$V$} q0 | q2;
      box {$U$} q1 | q2;
      }(q2,q1,q0);
    box {$M$} q2;
  \end{yquant}
\end{tikzpicture}}
\caption{The quantum circuit that transforms the quantum state $\ket{\F}$ represented by the LimTDD into $\ket{000}$. The two dotted boxes correspond to the processing of the $y_1$ and $y_1'$ nodes, respectively.
In this circuit $U = \frac{1}{\sqrt{2}}\left[\begin{smallmatrix} 
		1 & 1\\ 
		-1 & 1\\
\end{smallmatrix}\right]$, $V = \frac{1}{\sqrt{3}}\left[\begin{smallmatrix} 
		\sqrt{2} & 1\\ 
		-1 & \sqrt{2}\\
\end{smallmatrix}\right]$, $W = \frac{1}{\sqrt{11}}\left[\begin{smallmatrix} 
		2\sqrt{2} & \sqrt{3}\\ 
		-\sqrt{3} & 2\sqrt{2}\\
\end{smallmatrix}\right]$, $M = \frac{1}{\sqrt{23}}\left[\begin{smallmatrix} 
		\sqrt{11} & \sqrt{12}\\ 
		-\sqrt{12} & \sqrt{11}\\
\end{smallmatrix}\right]$.
}
\label{fig:pre_cir_noa}
\end{figure*}

We now provide a concrete example to illustrate the procedure of our algorithm, with the quantum state to be prepared being $ \frac{2}{\sqrt{23}} (\ket{000} + \ket{001} + \frac{1}{\sqrt{2}} \ket{010} + \frac{i}{2} \ket{011} - \ket{100} - \frac{1}{\sqrt{2}} \ket{101} + \frac{1}{\sqrt{2}} \ket{110} + \ket{111}) $. A step-by-step demonstration of our algorithm on the LimTDD is given in Fig. \ref{fig:limtdd}.  The resulting quantum circuit for preparing the desired quantum state is shown in Fig. \ref{fig:pre_cir_noa}.

At first, the quantum state represented by the LimTDD is
    $ \frac{2}{\sqrt{23}} Z \otimes I \otimes I \ket{v_{20}} $
    where:
    \begin{eqnarray*}
         \ket{v_{20}} &=& \ket{0} \ket{v_{10}} + \ket{1} (Z \otimes I \ket{v_{11}}) \\
     \ket{v_{10}} &=& \ket{0} \ket{v_{00}} + \frac{1}{\sqrt{2}} \ket{1} (S \ket{v_{01}}) \\
     \ket{v_{11}} &=& \ket{0} \ket{v_{01}} + \ket{1} (X \ket{v_{01}}) \\
     \ket{v_{00}} &=& \ket{0} + \ket{1} \\
     \ket{v_{01}} &=& \ket{0} + \frac{1}{\sqrt{2}} \ket{1}.
    \end{eqnarray*}

We then explain the procedure of the algorithm step-by-step:

\begin{enumerate}

    \item \textbf{Cancel the Operator on the Incoming Edge}:
    \begin{itemize}
        \item Apply a $Z$ gate on qubit $ q_2 $ to cancel the $ Z \otimes I \otimes I $ operator on the incoming edge of the LimTDD. The state becomes:
        $\ket{v_{20}} $.
    \end{itemize}

    \item \textbf{Process the $ v_{20} $ Node}:
    \begin{enumerate} 
        
    \item \textbf{Process the High-Edge Operator of $ v_{20} $ Node}:
    \begin{itemize}
        \item Apply a CZ gate with $ q_2 $ as controls and $ q_1 $ as the target to cancel the $ Z \otimes I $ operator on the high-edge of the $v_{20}$ node. The state becomes:
    $ \ket{0} \ket{v_{00}} + \ket{1} \ket{v_{11}}$.
    \end{itemize}

    \item \textbf{Process the $ v_{10} $ Node}:
    \begin{itemize}
        \item \textbf{Process the High-Edge Operator of $ v_{10} $ Node}:
        \begin{itemize}
            \item Apply a controlled-controlled-$S^\dag$ gate with control condition $\ket{0}_2\ket{1}_1$ and target qubit $q_0$ to cancel the $S$ operator on the high-edge of the $ y_1 $ node. The state becomes:
            $ \ket{0} (\ket{0}\ket{v_{00}} + \frac{1}{\sqrt{2}} \ket{1}\ket{v_{01}}) + \ket{1} \ket{v_{11}} $.
        \end{itemize}
    
        \item \textbf{Process the outgoing weights of $ v_{00}$ and $v_{01}$ Nodes with prefix path 00 and 01}:
        \begin{itemize}
            \item Since $ \ket{v_{00}} = \ket{0} + \ket{1} $, apply a controlled-$U$ gate with control condition $\ket{0}_2\ket{0}_1$ transform the state to:
            $ \ket{0} (\sqrt{2}\ket{0}\ket{0} + \frac{1}{\sqrt{2}} \ket{1}\ket{v_{01}}) + \ket{1} \ket{v_{11}} $
            where $ U = \frac{1}{\sqrt{2}} \begin{bmatrix} 1 & 1 \\ -1 & 1 \end{bmatrix} $.
            \item Since $ \ket{v_{01}} = \ket{0} + \frac{1}{\sqrt{2}}\ket{1} $, apply a controlled-V gate with control condition $\ket{0}_2\ket{1}_1$ transform the state to:
            $\ket{0} (\sqrt{2}\ket{0} + \frac{\sqrt{3}}{2} \ket{1})\ket{0} + \ket{1} \ket{v_{11}} $
            where $ V = \frac{\sqrt{2}}{\sqrt{3}}\left[\begin{array}{cccc} 
		  1 & \frac{1}{\sqrt{2}}\\ 
		  -\frac{1}{\sqrt{2}} & 1\\
            \end{array}\right]$.
        \end{itemize}

        \item \textbf{Adjust the Weights on Outgoing Edges of the $v_{10}$ Node}:
        \begin{itemize}
            \item Apply a controlled-$W$ gate with control condition $\ket{0}_2$  to adjust the weights on the outgoing edges of the $ v_{10} $ node and change the state to: $\frac{\sqrt{11}}{2}\ket{0}\ket{0}\ket{0} + \ket{1} \ket{v_{11}} $
            where $ W = \frac{1}{\sqrt{11}}\left[\begin{smallmatrix} 
		  2\sqrt{2} & \sqrt{3}\\ 
		  -\sqrt{3} & 2\sqrt{2}\\
        \end{smallmatrix}\right]$.
        \end{itemize}
    
    \end{itemize}

    \item \textbf{Process the $ v_{11} $ Node}:
    \begin{itemize}
        \item \textbf{Process the High-Edge Operator of $ v_{11} $ Node}:
        \begin{itemize}
            \item Apply a CCX gate with control condition $\ket{1}_2\ket{1}_1$ to cancel the operator on the high-edge of the $v_{11}$ node. The state becomes:
            $ \frac{\sqrt{11}}{2} \ket{0} \ket{0} \ket{0}  + \sqrt{3} \ket{1} (\ket{0}+\ket{1}) \ket{v_{01}} $.
        \end{itemize}

        \item \textbf{Process the outgoing weights of $v_{01}$ Node with prefix path 11}:
        \begin{itemize}
            \item Apply a controlled-$V$ gate with $\ket{1}_2$, the state will be changed to 
            $\frac{\sqrt{11}}{2} \ket{0}\ket{0}\ket{0}+\ket{1}\big(\frac{\sqrt{3}}{\sqrt{2}}\ket{0}+\frac{\sqrt{3}}{\sqrt{2}}\ket{1}\big)\ket{0}$.
        \end{itemize}

        \item \textbf{ Adjust the Weights on Outgoing Edges of the $v_{11}$ Node}:
        \begin{itemize}
            \item Apply a further gate $U$ controlled by $\ket{1}_2$ and change the state to 
            $\frac{\sqrt{11}}{2} \ket{0}\ket{0}\ket{0}+\sqrt{3}\ket{1}\ket{0}\ket{0}$.
        \end{itemize}
    \end{itemize}
    
    \item \textbf{Adjust the Weights on Outgoing Edges of the $y_2$ Node}:
    \begin{itemize}
        \item Use a $M = \frac{1}{\sqrt{23}}\left[\begin{smallmatrix} 
		\sqrt{11} & \sqrt{12}\\ 
		-\sqrt{12} & \sqrt{11}\\
        \end{smallmatrix}\right]$ gate to adjust the weights on the outgoing edges of the $v_{20}$ node, the state becomes: 
        $ \frac{\sqrt{23}}{2}  \ket{0} \ket{0} \ket{0} $. Note that the coefficient $\frac{\sqrt{23}}{2} $ is cancelled with the ignored coefficient $\frac{2}{\sqrt{23}}$.
    \end{itemize}

    \end{enumerate}
\end{enumerate}

\subsection{Complexity}

In this subsection, we consider the time complexity and gate complexity of the algorithm. Here, the time complexity is determined by the number of recursive calls to the function $\textsc{StatePre1}$, and the gate complexity is determined by the number of gates in the returned circuit.

\subsubsection{Time Complexity}

During the process, we traverse the decision graph in a depth-first manner, so each reduced path is visited exactly once. Given there are \( p \) reduced paths in the LimTDD, and each path contains \( n \) non-terminal nodes (where \( n \) is the number of qubits), the time complexity is \( \mathcal{O}(np) \). 

To optimise this, we can introduce a table to cache the results of previously computed nodes. This allows us to reuse these results if a node is encountered multiple times, thereby reducing the time complexity to \( \mathcal{O}(m) \), where \( m \) is the total number of nodes in the LimTDD.

\subsubsection{Gate Complexity}

The gate complexity is predominantly influenced by the step of cancelling all high-edge operators and outgoing weights using multi-qubit controlled quantum gates. For a node at level \( k \in \{2, \ldots, n\} \) (with the terminal node defined as level 0), the high-edge operator has the form \( O_{k-1} \otimes \cdots \otimes O_1 \), which includes at most \( k-1 \) non-trivial (i.e., non-identity) single-qubit operators. This requires \( n+2-k \) qubit-controlled gates for elimination. Additionally, the outgoing weights of a node at level \( k \in \{1, \ldots, n\} \) need to be reduced using \( n+1-k \) qubit-controlled gates. On any given path, there can be at most \( n \) non-trivial outgoing weights (i.e., weights not of the form \([w, 0]\) for some complex number \( w \neq 0 \)) that need to be reduced. Moreover, up to \( n \) single-qubit gates are required to eliminate the operator on the incoming edge of the LimTDD.

Consequently, the upper bound on the gate complexity is:
\begin{itemize}
    \item \( p(n+2-s) \) \( s \)-qubit gates for \( s \in \{2, \ldots, n\} \),
    \item \( n+1 \) single-qubit gates.
\end{itemize}

It is worth noting that the \( s \)-qubit gates may involve fewer qubits, as they do not always need to be controlled by all the states corresponding to their prefix path.

\subsubsection{Special Case}

In the special case where the decision diagram is in tower form (i.e., the 0- and 1-successors of each non-terminal node are identical), there is only one reduced path. In this scenario, the upper bound on the gate complexity reduces to \( 2n \) single-qubit gates and \( \frac{n(n-1)}{2} \) two-qubit gates. In such cases, this algorithm is optimal compared to the other algorithms proposed below.

\subsection{Further optimisation}

In this subsection, we explore strategies to reduce the number of multi-qubit control gates by introducing an ancilla qubit \( q_{a} \). 

Consider the scenario where we need to eliminate an operator \( O_j \otimes \cdots \otimes O_1 \) on the high edge of a node \( v \). The step typically involves using a multi-qubit gate to eliminate all operators \( O_i \) for \( i \in \{1, \ldots, j\} \) with the same control qubits. Suppose the control condition is \( \ket{k} \), which is an \( s \geq 2 \) qubit state, this would normally require \( j \) \( s+1 \)-qubit gates. However, with an ancilla qubit \( q_{a} \) available, we can proceed as follows:
\begin{itemize}
    \item Apply an \( MCX \) gate with control condition \( \ket{k} \) and target \( q_{a} \) to mark the state \( \ket{k} \) as open, resulting in the state \( \ket{1}_{a'}\ket{k}\ket{v} + \ket{0}_{a}\ket{\rm Res} \).
    \item Use \( \ket{1}_{a} \) as the control condition to eliminate \( O_j \otimes \cdots \otimes O_1 \).
    \item Apply another \( MCX \) gate with control condition \( \ket{k} \) and target \( q_{a} \) to recover \( q_{a} \).
\end{itemize}
This approach reduces the number of \( s+1 \)-qubit gates from \( j \) to 2, while introducing \( j \) 2-qubit gates.

Note that this strategy is applicable to all four algorithms proposed in this paper, and has no relations with the Alg. \ref{alg:State_Pre_onea} introduced below.

\section{LimTDD based QSP with One Ancilla Qubits} \label{sec:Algorithms_onea}


In this section, we present a quantum state preparation algorithm leveraging a single ancilla qubit. Our method is conceptually inspired by the algorithm by Mozafari et al. \cite{mozafari_efficient_2022}, which introduced an efficient algorithm for QSP using decision diagrams. Their method begins with the \(\ket{0}\) state and incrementally constructs the target quantum state by traversing the decision diagram path by path. A key feature of their algorithm is the use of an ancilla qubit to mark paths that have already been processed. By leveraging this ancilla qubit as a control qubit, subsequent preparation steps do not interfere with previously prepared paths. Consequently, both the time complexity and the gate complexity of the resulting quantum circuit scale with the number of reduced paths in the decision diagram.

\begin{figure}[htbp]
    \centering

    \resizebox{0.2\textwidth}{!}{




\begin{tikzpicture}[
>=latex,
line join=bevel,
every node/.style={minimum size=1.4cm, inner sep=0pt,thick, font=\Huge}, 
every path/.style={line width=1.5pt} 
]


  \node (i) at (138.67bp,21.667bp) [draw=red,circle,font=\LARGE] {$\frac{i}{2}$};
\node (1) at (21.667bp,21.667bp) [draw=red,circle,font=\LARGE] {$1$};
  \node (-1) at (229.67bp,21.667bp) [draw=red,circle,font=\LARGE] {$-1$};
  \node (2) at (82.667bp,21.667bp) [draw=red,circle,font=\LARGE] {$\frac{1}{\sqrt{2}}$};
  \node (-2) at (183.67bp,21.667bp) [draw=red,circle,font=\LARGE] {$-\frac{1}{\sqrt{2}}$};
  \node (y01) at (120.67bp,100.49bp) [draw=red,circle] {$y_0$};
  \node (y02) at (180.67bp,100.49bp) [draw=red,circle] {$y_0$};
  \node (y03) at (60.667bp,100.49bp) [draw=red,circle] {$y_0$};
  \node (y10) at (70.667bp,178.8bp) [draw=red,circle] {$y_1$};
  \node (y11) at (130.67bp,178.8bp) [draw=red,circle] {$y_1$};
  \node (y2) at (100.67bp,257.11bp) [draw=red,circle] {$y_2$};
  \node (-0) at (100.67bp,332.26bp) [draw,draw=none] {};
  \draw [red,->,dotted] (y01) ..controls (107.14bp,72.152bp) and (101.89bp,61.528bp)  .. (2);
  \draw [blue,->] (y01) ..controls (128.94bp,71.853bp) and (131.7bp,62.632bp)  .. (i);
  \draw [red,->,dotted] (y02) ..controls (209.22bp,72.102bp) and (212.18bp,62.628bp)  .. (-1);
  \draw [blue,->] (y02) ..controls (177.5bp,72.152bp) and (172.38bp,61.528bp)  .. (-2);
  \draw [blue,->] (y03) ..controls (46.958bp,72.485bp) and (41.488bp,61.71bp)  .. (1);
  \draw [red,->,dotted] (y03) ..controls (68.581bp,71.853bp) and (71.221bp,62.632bp)  .. (2);
  \draw [red,->,dotted] (y10) ..controls (47.918bp,151.69bp) and (37.26bp,136.79bp)  .. (31.667bp,121.64bp) .. controls (23.786bp,100.31bp) and (21.367bp,74.739bp)  .. (1);
  \draw [blue,->] (y10) ..controls (88.277bp,150.92bp) and (96.192bp,138.84bp)  .. (y01);
  \draw [red,->,dotted] (y11) ..controls (148.28bp,150.92bp) and (156.19bp,138.84bp)  .. (y02);
  \draw [blue,->] (y11) ..controls (107.03bp,152.03bp) and (93.66bp,137.46bp)  .. (y03);
  \draw [red,->,dotted] (y2) ..controls (89.863bp,228.63bp) and (86.014bp,218.84bp)  .. (y10);
  \draw [blue,->] (y2) ..controls (111.47bp,228.63bp) and (115.32bp,218.84bp)  .. (y11);
  \draw [red,->] (-0) ..controls (100.67bp,307.06bp) and (100.67bp,298.38bp)  .. (y2);
\end{tikzpicture}

%
%
}

    \caption{An example of Multiple-terminal ADD \cite{mozafari_efficient_2022} representing the quantum state     $\frac{2}{\sqrt{23}}[1,1,\frac{1}{\sqrt{2}},\frac{i}{2},-1,-\frac{1}{\sqrt{2}},\frac{1}{\sqrt{2}},1]^T$. The coefficient $\frac{2}{\sqrt{23}}$ was omitted here.
    }
    \label{fig:add}
\end{figure}
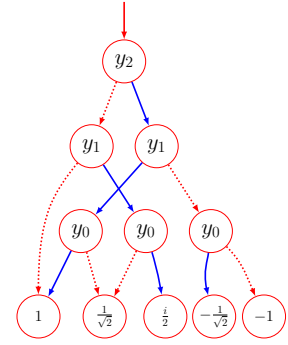

However, the decision diagram employed in \cite{mozafari_efficient_2022} - the multi-terminal Algebraic Decision Diagram (ADD)- is less compact compared to the LimTDD used in this paper. As an example, Fig.~\ref{fig:add} illustrates the ADD representation of the quantum state presented in Example~\ref{exp:q_state}. While the ADD comprises 7 reduced paths, the corresponding LimTDD representation contains only 3, demonstrating a substantial reduction in path counts. Motivated by the observations above, we propose a LimTDD-based QSP algorithm that utilises a single ancilla qubit. 

Although our algorithm draws inspiration from \cite{mozafari_efficient_2022}, their implementation details differ substantially, owing to the fundamental structural distinctions between ADDs and LimTDDs. Specifically, our algorithm eliminates the need to compute the accumulated probability along different branches, thereby streamlining the process. Instead of marking paths, we repurpose the ancilla qubit to indicate the open or closed status of a node. Additionally, we design a circuit that transforms the quantum state represented by the LimTDD into \(\ket{0}\), which is the opposite of the reverse transformation presented in \cite{mozafari_efficient_2022}.

In the following subsections, we provide a detailed exposition of our proposed algorithm.

\subsection{Algorithm}

The core of this algorithm involves introducing an ancilla qubit to control the opening and closing of nodes within the LimTDD. By ensuring that the current node (and its associated subtrees) is the only open node at any moment, we can process it without affecting other nodes. This approach repurposes the ancilla qubit as a control condition, thereby reducing the number of control qubits required in Alg.~\ref{alg:State_Pre_noa}.

Our algorithm, detailed in Alg.~\ref{alg:State_Pre_onea}, begins by eliminating the operator on the incoming edge of the root node. It then proceeds recursively: for each node, we first cancel the operator on its high-edge, process its 0-successor, followed by its 1-successor, and finally adjust the weights on the outgoing edges of the current node.

Throughout this procedure, an ancilla qubit is used to mark the current node \( v \) being processed, effectively indicating the open part of the decision diagram. Initially, the entire LimTDD (the root node) is open, marked by $\ket{1}_a$. When a branch node is encountered, we first use a multi-controlled-\( X \) ($MCX$) gate to close its 1-successor, with the control condition set to the branch condition of \( \high(v) \). We then proceed to process the 0-successor. 

Subsequently, we toggle the open/close condition of the two successors using the branch condition of \( v \), and process the 1-successor. Finally, we reopen the 0-successor using the branch condition of \( \low(v) \), ensuring the original node remains open. This process is applied recursively; upon completion, the root node is left open, thus the ancilla qubit has been successfully returned to the \( \ket{1}_a \) state.

To illustrate, consider a node \( v \) with the form
\[
\Circled{v_0} \overset{I}{\dashleftarrow} \Circled{v} \xrightarrow{\lambda O} \Circled{v_1},
\]
marked open by \( q_a \) with branch condition \( \ket{p} \). The state can be represented as
$
\ket{1}_a \ket{p} \ket{v} + \ket{0}_a \ket{\rm Res} = \ket{1}_a \ket{p} \ket{0}_v \ket{v_0} + \lambda \ket{1}_a \ket{p} \ket{1}_v (O \ket{v_1}) + \ket{0}_a \ket{\rm Res},
$
where \( \ket{\rm Res} \) is orthogonal to \( \ket{p} \ket{*} \). The operator \( O \) can be eliminated using a controlled \( O^\dag \) operator with control condition \( \ket{1}_a \ket{1}_v \), transforming the state to
\[
\ket{1}_a \ket{p} \ket{0}_v \ket{v_0} + \lambda \ket{1}_a \ket{p} \ket{1}_v \ket{v_1} + \ket{0}_a \ket{\rm Res}.
\]
A $MCX$ gate with control condition \( \ket{p} \ket{1}_v \) is then applied to close the \( v_1 \) node, resulting in the state
\[
\ket{1}_a \ket{p} \ket{0}_v \ket{v_0} + \lambda \ket{0}_a \ket{p} \ket{1}_v \ket{v_1} + \ket{0}_a \ket{\rm Res}.
\]
Using \( \ket{1}_a \) as the control qubit, we process the \( v_0 \) node, obtaining a circuit \( cir_0 \) that transforms \( \ket{1}_a \ket{v_0} \) to \( ||v_0|| \ket{1}_a \ket{0}^{\otimes k} \) for some \( k \), while leaving other parts unchanged. Applying this circuit, the original state becomes
\[
||v_0|| \ket{1}_a \ket{p} \ket{0}_v \ket{0}^{\otimes k} + \lambda \ket{0}_a \ket{p} \ket{1}_v \ket{v_1} + \ket{0}_a \ket{\rm Res}.
\]
A $MCX$ gate with control condition \( \ket{p} \) is then applied to \( q_a \), changing the state to
\[
||v_0|| \ket{0}_a \ket{p} \ket{0}_v \ket{0}^{\otimes k} + \lambda \ket{1}_a \ket{p} \ket{1}_v \ket{v_1} + \ket{0}_a \ket{\rm Res}.
\]
We then process the \( v_1 \) node with control condition \( \ket{1}_a \), transforming the state to
\[
||v_0|| \ket{0}_a \ket{p} \ket{0}_v \ket{0}^{\otimes k} + \lambda ||v_1|| \ket{1}_a \ket{p} \ket{1}_v \ket{0}^{\otimes k} + \ket{0}_a \ket{\rm Res}.
\]
Finally, a $MCX$ gate with control condition \( \ket{p} \ket{0}_v \) reopens the \( v_0 \) part, resulting in the state
\[
\ket{1}_a \ket{p} (||v_0|| \ket{0}_v + \lambda ||v_1|| \ket{1}_v) \ket{0}^{\otimes k} + \ket{0}_a \ket{\rm Res}.
\]
An operator
\[
\frac{1}{\sqrt{1 + |c|^2}}
\begin{bmatrix}
1 & c^\dag \\
-c & 1 \\
\end{bmatrix}
\]
with \( c = \lambda \cdot ||v_1|| / ||v_0|| \) and control condition \( \ket{1}_a \) is then applied, transforming the state to
\[
||v|| \ket{1}_a \ket{p} \ket{0}^{\otimes k+1} + \ket{0}_a \ket{\rm Res},
\]
completing the process of the $v$ node.

\begin{algorithm}[htbp]
\caption{$\textsc{StatePre2}(v, q_a, p)$}
\begin{algorithmic}[1]
\Require{A node $v$ of a LimTDD $\F$ representing an $(n-|p|)$-qubit quantum state $\ket{\psi}$; an ancilla qubit $q_a$ which marked $v$ open under the $|p|$-bit branch condition $p$. For root node, the branch condition is empty.}
\Ensure{A quantum circuit $C$ with unitary $U$ such that $U \ket{1}_a\ket{p}\ket{\psi} = \ket{1}_a\ket{p}\ket{0}^{\otimes {n-|p|}}$.}
\vspace{0.4em}

\State $C \gets \mathrm{QuantumCircuit}(n+1)$ {\color{gray}\Comment{Empty quantum circuit}}
\If{$v$ is the terminal node}
\State $||v|| \gets 1$
\State \Return $C$
\EndIf
\State Let $\w\big((v,\high(v))\big)=\lambda \cdot O$. Add to $C$ a $\ket{1}_a\ket{1}_v$-controlled $O^\dag$ gate  \quad {\color{gray}\Comment{Apply BC2 on $v$ }} 
\If{$\low(v)=\high(v)$}
\State $C_0 \gets \textsc{StatePre2}\big(\low(v),q_a,p\big)$
\State Append $C_0$ to $C$
\Else
\State Add to $C$ a $\ket{p}\ket{1}_v$-controlled $X$ gate on $q_a$ to close the high-branch of $v$
\State $C_0 \gets \textsc{StatePre2}\big(\low(v),q_a,p0\big)$
\State Append $C_0$ to $C$
\State Append to $C$ a $\ket{p}$-controlled $X$ gate on $q_a$ to close the low-branch and open the high-branch of $v$
\State $C_1 \gets \textsc{StatePre2}\big(\high(v),q_a,p1\big)$
\State Append $C_1$ to $C$
\State Append to $C$ a $\ket{p}\ket{0}_v$-controlled $X$ gate on $q_a$ to open the low-branch of $v$
\EndIf
\State $c \gets \lambda \cdot ||\high(v)||/||\low(v)||$
\State Append to $C$ a $\ket{1}_a$-controlled $R(c)$ gate on $q_v$, the qubit corresponds to $v$ in $\F$ \quad {\color{gray}\Comment{Apply BC3 on $v$}}
\vspace{0.2em}
\State $||v||\gets \sqrt{||\low(v)||^2+|\lambda|^2 \cdot ||\high(v)||^2}$ 
\State \Return $C$
\end{algorithmic}
\label{alg:State_Pre_onea} 
\end{algorithm}

\subsection{An Example}\label{sec:exp}

\begin{figure*}[htbp]
\centering

\subfigure[]{
    \resizebox{0.2\textwidth}{!}{




\begin{tikzpicture}[
>=latex,
line join=bevel,
every node/.style={minimum size=1.4cm, inner sep=0pt,thick, font=\LARGE}, 
every path/.style={line width=1.5pt} 
]
\node (1) at (128.67bp,18.0bp) [draw=red,circle,font=\Huge] {$1$};
  \node (y01) at (46.668bp,108.9bp) [draw=red,circle,font=\Huge] {$v_{00}$};
  \node (y02) at (185.67bp,108.9bp) [draw=red,circle,font=\Huge] {$v_{01}$};
  \node (y10) at (66.668bp,202.96bp) [draw=red,circle,font=\Huge] {$v_{10}$};
  \node (y11) at (165.67bp,202.96bp) [draw=red,circle,font=\Huge] {$v_{11}$};
  \node (y2) at (107.67bp,297.02bp) [draw=red,circle,font=\Huge] {$v_{20}$};
  \node (-0) at (107.67bp,387.93bp) [draw,draw=none] {};
  \draw [red,->,dotted] (y01) ..controls (29.551bp,79.314bp) and (25.128bp,64.774bp)  .. (32.668bp,54.0bp) .. controls (47.583bp,32.686bp) and (76.911bp,24.306bp)  .. (1);
  \definecolor{strokecol}{rgb}{0.0,0.0,0.0};
  \pgfsetstrokecolor{strokecol}
  \draw (53.668bp,61.875bp) node {};
  \draw [blue,->] (y01) ..controls (73.917bp,78.36bp) and (94.182bp,56.39bp)  .. (1);
  \draw (116.67bp,61.875bp) node {};
  \draw [red,->,dotted] (y02) ..controls (166.06bp,77.316bp) and (153.94bp,58.425bp)  .. (1);
  \draw (180.67bp,61.875bp) node {};
  \draw [blue,->] (y02) ..controls (205.77bp,80.723bp) and (212.16bp,65.699bp)  .. (204.67bp,54.0bp) .. controls (194.36bp,37.899bp) and (174.56bp,29.102bp)  .. (1);
  \draw (230.29bp,61.875bp) node {$\frac{1}{\sqrt{2}}$};
  \draw [red,->,dotted] (y10) ..controls (31.234bp,190.19bp) and (11.932bp,180.12bp)  .. (2.6676bp,163.81bp) .. controls (-5.0407bp,150.24bp) and (6.2358bp,136.64bp)  .. (y01);
  \draw (23.668bp,155.93bp) node {};
  \draw [blue,->] (y10) ..controls (59.707bp,169.93bp) and (56.313bp,154.3bp)  .. (y02);
  \draw (105.543bp,155.93bp) node {$\frac{1}{\sqrt{2}} S$};
  \draw [red,->,dotted] (y11) ..controls (163.61bp,171.29bp) and (164.02bp,158.88bp)  .. (166.67bp,148.06bp) .. controls (167.5bp,144.65bp) and (168.68bp,141.19bp)  .. (y02);
  \draw (187.67bp,155.93bp) node {};
  \draw [blue,->] (y11) ..controls (193.01bp,184.7bp) and (203.51bp,175.28bp)  .. (208.67bp,163.81bp) .. controls (212.85bp,154.5bp) and (210.23bp,144.22bp)  .. (y02);
  \draw (230.42bp,155.93bp) node {$X$};
  \draw [red,->,dotted] (y2) ..controls (93.74bp,264.75bp) and (86.177bp,247.77bp)  .. (y10);
  \draw (110.67bp,249.99bp) node {};
  \draw [blue,->] (y2) ..controls (125.36bp,273.0bp) and (130.96bp,265.2bp)  .. (135.67bp,257.87bp) .. controls (140.87bp,249.76bp) and (146.13bp,240.69bp)  .. (y11);
  \draw (172.92bp,249.99bp) node {$Z\otimes I$};
  \draw [red,->] (-0) ..controls (107.67bp,358.85bp) and (107.67bp,343.46bp)  .. (y2);
  \draw (154.54bp,344.05bp) node {$\frac{2}{\sqrt{23}} Z\otimes I\otimes I$};
\end{tikzpicture}

%
%
}
}
\subfigure[]{
\begin{tikzpicture}
  \begin{yquant}[register/minimum height=12mm, operator/separation=2mm, control style={radius=2pt}, subcircuit box style={dashed}]
    qubit {$\ket{1}_a$} a;
    qubit {$\ket{q_2}$} q2;
    qubit {$\ket{q_1}$} q1;
    qubit {$\ket{q_0}$} q0;
    box {$Z$} q2;
    z q1 | q2, a;
    box {$X$} a | q2;
    subcircuit {
      qubit {} a;
      qubit {} q2;
      qubit {} q1;
      qubit {} q0;
      
      box {$S^\dag$} q0 | q1, a;
      box {$X$} a | q1, ~q2;
      box {$U$} q0 | a;
      box {$X$} a | ~q2;
      box {$V$} q1 | a;
      box {$X$} a | ~q1, q2;
      box {$W$} q1 | a;
    } (a,q2,q1,q0);
    x a;
    subcircuit {
      qubit {} a;
      qubit {} q2;
      qubit {} q1;
      qubit {} q0;
      x q0 | q1,a;
      box {$V$} q0 | a;
      box {$U$} q1 | a;
      }(a,q2,q1,q0);
    x a | ~q2;
    box {$M$} q2 | a;
  \end{yquant}
\end{tikzpicture}}
\caption{The quantum circuit that transforms the quantum state $\ket{1}\ket{\F}$ into $\ket{1}\ket{000}$. The two dotted boxes correspond to the processing of the $y_1$ and $y_1'$ nodes, respectively.
In this circuit $U = \frac{1}{\sqrt{2}}\left[\begin{smallmatrix} 
		1 & 1\\ 
		-1 & 1\\
\end{smallmatrix}\right]$, $V = \frac{\sqrt{2}}{\sqrt{3}}\left[\begin{smallmatrix} 
		1 & \frac{1}{\sqrt{2}}\\ 
		-\frac{1}{\sqrt{2}} & 1\\
\end{smallmatrix}\right]$, $W = \frac{1}{\sqrt{11}}\left[\begin{smallmatrix} 
		2\sqrt{2} & \sqrt{3}\\ 
		-\sqrt{3} & 2\sqrt{2}\\
\end{smallmatrix}\right]$, $M = \frac{1}{\sqrt{23}}\left[\begin{smallmatrix} 
		\sqrt{11} & \sqrt{12}\\ 
		-\sqrt{12} & \sqrt{11}\\
\end{smallmatrix}\right]$.
}
\label{fig:pre_cir_onea}
\end{figure*}

In this subsection, we will also use the LimTDD given in Fig. \ref{fig:limtdd} as an example to provide a step-by-step illustration of the algorithm.

We start by adding an ancilla qubit $q_a$ and the initial state  becomes $ \frac{2}{\sqrt{23}} \ket{1}_a (Z \otimes I \otimes I \ket{v_{20}}) $.

Then we explain the procedure of the algorithm step-by-step.

\begin{enumerate}

    \item \textbf{Cancel the Operator on the Incoming Edge}:
    \begin{itemize}
        \item Apply a $Z$ gate on qubit $ q_2 $ to cancel the $ Z \otimes I \otimes I $ operator on the incoming edge of the LimTDD. The state becomes:
        $ \ket{1}_a \ket{v_{20}} $.
    \end{itemize}

    \item \textbf{Process the $ v_{20} $ Node}:
    \begin{enumerate} 
        
    \item \textbf{Process the High-Edge Operator of $ v_{20} $ Node}:
    \begin{itemize}
        \item Apply a CCZ gate with $ q_a $ and $ q_2 $ as controls and $ q_1 $ as the target to cancel the $ Z \otimes I $ operator on the high-edge of the $ v_{20} $ node. The state becomes:
        $ \ket{1}_a (\ket{0} \ket{v_{10}} + \ket{1} \ket{v_{11}}) $.
    \end{itemize}

    \item \textbf{Process the $ v_{10} $ Node}:
    \begin{itemize}
        \item \textbf{Process the High-Edge Operator of $ v_{10} $ Node}:
        \begin{itemize}
            \item Use a $CX$ gate with $ q_2 $ as the control qubit and $ q_a $ as the target qubit to close the $ v_{11} $ node. The state becomes:
            $ \ket{1}_a \ket{0} \ket{v_{10}} + \ket{0}_a \ket{1} \ket{v_{11}} $.
            \item Apply a Controlled-Controlled-$S^\dag$ gate with $ q_a $ and $ q_1 $ as controls and $ q_0 $ as the target to cancel the $Z$ operator on the high-edge of the $ v_{10} $ node. The state becomes:
            $ \ket{1}_a \ket{0} (\ket{0}\ket{v_{00}} + \frac{1}{\sqrt{2}} \ket{1} \ket{v_{01}}) + \ket{0}_a\ket{1} \ket{v_{11}} $.
        \end{itemize}
    
        \item \textbf{Process the outgoing weights of $ v_{00}$ and $v_{01}$ Nodes with prefix path 00 and 01}:
        \begin{itemize}
            \item Use a CCX gate with control condition $\ket{0}_2\ket{1}_1$ to close the $v_{01}$ node, and the state becomes: $ \ket{1}_a \ket{0} \ket{0}\ket{v_{00}} + \frac{1}{\sqrt{2}} \ket{0}_a \ket{0}\ket{1} \ket{v_{01}} + \ket{0}_a\ket{1} \ket{v_{11}} $.
            \item Apply a controlled-U gate with $ q_a $ as the control qubit to transform the state to:
            $ \sqrt{2}\ket{1}_a \ket{0} \ket{0}\ket{0} + \frac{1}{\sqrt{2}} \ket{0}_a \ket{0}\ket{1} \ket{v_{01}} + \ket{0}_a\ket{1} \ket{v_{11}} $.
            \item Apply a CX gate with control condition $\ket{0}_2$ and target qubit $q_a$ to close the $v_{00}$ node and open the $v_{01}$ node and change the state to: $ \sqrt{2}\ket{0}_a \ket{0} \ket{0}\ket{0} + \frac{1}{\sqrt{2}} \ket{1}_a \ket{0}\ket{1} \ket{v_{01}} + \ket{0}_a\ket{1} \ket{v_{11}} $.
            \item Apply a controlled-V gate with $ q_a $ as the control qubit to transform the state to:
            $ \sqrt{2}\ket{0}_a \ket{0} \ket{0}\ket{0} + \frac{\sqrt{3}}{2} \ket{1}_a \ket{0}\ket{1} \ket{0} + \ket{0}_a\ket{1} \ket{v_{11}} $.
            \item Apply a CCX gate with control condition $\ket{0}_2\ket{0}_1$ to reopen the $v_{00}$ node and change the state to: $ \ket{1}_a \ket{0} (\sqrt{2}\ket{0}+\frac{\sqrt{3}}{2}\ket{1})\ket{0} + \ket{0}_a\ket{1} \ket{v_{11}} $.
        \end{itemize}

        \item \textbf{Adjust the Weights on Outgoing Edges of the $v_{10}$ Node}:
        \begin{itemize}
            \item Apply a controlled-$W$ gate with $ q_a $ as the control qubit to adjust the weights on the outgoing edges of the $ v_{10} $ node and change the state to: $ \frac{\sqrt{11}}{2}\ket{1}_a \ket{0} \ket{0}\ket{0} + \ket{0}_a\ket{1} \ket{v_{11}} $.
        \end{itemize}
    
    \end{itemize}

    \item \textbf{Process the $ v_{11} $ Node}:
    \begin{itemize}
        \item \textbf{Process the High-Edge Operator of $ v_{11} $ Node}:
        \begin{itemize}
            \item Use an $X$ gate on $ q_a $ to switch the branches and open the $ v_{11} $ node. The state becomes:
            $ \frac{\sqrt{11}}{2}\ket{0}_a \ket{0} \ket{0}\ket{0} + \ket{1}_a\ket{1} \ket{v_{11}} $.
            \item Apply a CCX gate controlled by $ q_a $ and $ q_1 $ to cancel the operator on the high-edge of the $ v_{11}$ node. The state becomes:
            $ \frac{\sqrt{11}}{2} \ket{0}_a \ket{0} \ket{0} \ket{0}  + \sqrt{3} \ket{1}_a \ket{1} (\ket{0}+\ket{1}) \ket{v_{01}} $.
        \end{itemize}

        \item \textbf{Process the outgoing weights of $v_{01}$ Node with prefix path 11}:
        \begin{itemize}
            \item Apply a controlled-$V$ gate with $ q_a $ as the control qubit, the state will be changed to 
            $\frac{\sqrt{11}}{2}\ket{0}_a\ket{0}\ket{0}\ket{0}+\ket{1}_a\ket{1}\big(\frac{\sqrt{3}}{\sqrt{2}}\ket{0}+\frac{\sqrt{3}}{\sqrt{2}}\ket{1}\big)\ket{0}$.
        \end{itemize}

        \item \textbf{ Adjust the Weights on Outgoing Edges of the $v_{11}$ Node}:
        \begin{itemize}
            \item Apply a further gate $U$ controlled by $\ket{1}_a$ and change the state to 
            $\frac{\sqrt{11}}{2}\ket{0}_a\ket{0}\ket{0}\ket{0}+\sqrt{3}\ket{1}_a\ket{1}\ket{0}\ket{0}$.
            \item Apply a CX gate with the control qubit $q_2$ set to be $\ket{0}$ and target qubit $q_a$, to reopen the $v_{10}$ node, thus making the all branches of $v_{20}$ node open, and the state becomes: $\ket{1}_a\big(\frac{\sqrt{11}}{2}\ket{0}+\sqrt{3}\ket{1}\big)\ket{0}\ket{0}$.
        \end{itemize}

    \end{itemize}
    
    \item \textbf{Adjust the Weights on Outgoing Edges of the $y_2$ Node}:
    \begin{itemize}
        \item Use a controlled-$M$ gate with $ q_a $ as the control qubit to adjust the weights on the outgoing edges of the $ v_{20} $ node, the state becomes: 
        $ \frac{\sqrt{23}}{2} \ket{1}_a\ket{0} \ket{0} \ket{0} $. Also, the coefficient $\frac{\sqrt{23}}{2}$ cancels with complex weight $\frac{2}{\sqrt{23}}$ on the incoming edge.
    \end{itemize}

    \end{enumerate}
\end{enumerate}

The resulted quantum circuit for preparing the desired quantum state is shown in Fig. \ref{fig:pre_cir_onea}.

\subsection{Complexity}

\subsubsection{Time Complexity}

In this algorithm, we traverse the decision graph in a depth-first manner, ensuring that each reduced path is visited exactly once. Given that there are \( p \) reduced paths, the time complexity of this algorithm is \( \mathcal{O}(np) \), where \( n \) is the number of qubits.

\subsubsection{Gate Complexity}

The gate complexity is determined by the number of operations required to eliminate high-edge operators and to adjust outgoing weights. Specifically, high-edge operators are eliminated using 3-qubit controlled gates, and outgoing weights are adjusted using 2-qubit controlled gates. Additionally, the operator on the incoming edge is handled using no more than \( n \) single-qubit gates. For each branch node at level \( k \in \{2, \ldots, n\} \), the algorithm requires one controlled gate to close its 1-branch, one controlled gate to flip the open/close status of both branches, and one controlled gate to reopen the 0-branch. Given \( n-k \) nodes preceding the current node on the path, this necessitates two controlled gates with at most \( n-k+2 \) qubits and one controlled gate with at most \( n-k+1 \) qubits.

In summary, the upper bounds are:
\begin{itemize}
    \item $2p$ $n$-qubit gates,
    \item $3p$ $s$-qubit gates for $s\in \{4,\cdots ,n-1\}$,
    \item $\frac{n(n-1)}{2}p+3p$ 3-qubit gates,
    \item $np+3p$ 2-qubit gates,
    \item $n+1$ single-qubit gates.
\end{itemize}

\subsubsection{Special Case}

For decision diagrams in the tower form, the upper bound reduces to \( n \) single-qubit gates, \( n \) two-qubit gates, and \( \frac{n(n-1)}{2} \) three-qubit gates. In this scenario, the circuit generated by Alg.~\ref{alg:State_Pre_onea} is similar to that of Alg.~\ref{alg:State_Pre_noa}, with the exception that all gates are controlled by \( \ket{1}_a \), except those used to eliminate the operator on the incoming edge of the LimTDD.

\section{LimTDD based QSP with Sufficient Number of Ancilla Qubits} \label{sec:Algorithms_mul}

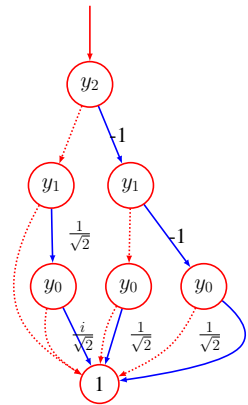
\begin{figure}[htbp]
    \centering

    \resizebox{0.2\textwidth}{!}{




\begin{tikzpicture}[
>=latex,
line join=bevel,
every node/.style={minimum size=1.4cm, inner sep=0pt,thick, font=\Huge}, 
every path/.style={line width=1.5pt} 
]

\pgfsetcolor{black}
  \pgfsetcolor{red}
  \draw [->,dotted] (63.708bp,276.95bp) .. controls (58.785bp,264.36bp) and (52.236bp,247.61bp)  .. (42.617bp,223.01bp);
  \definecolor{strokecol}{rgb}{0.0,0.0,0.0};
  \pgfsetstrokecolor{strokecol}
  \draw (58.042bp,249.99bp) node {};
  \pgfsetcolor{blue}
  \draw [->] (79.041bp,276.95bp) .. controls (84.238bp,264.36bp) and (91.15bp,247.61bp)  .. (101.3bp,223.01bp);
  \definecolor{strokecol}{rgb}{0.0,0.0,0.0};
  \pgfsetstrokecolor{strokecol}
  \draw (97.792bp,249.99bp) node {-1};
  \pgfsetcolor{red}
  \draw [->,dotted] (24.038bp,184.47bp) .. controls (11.17bp,161.99bp) and (-6.9875bp,121.8bp)  .. (3.4172bp,87.75bp) .. controls (11.048bp,62.778bp) and (34.929bp,43.932bp)  .. (64.047bp,26.661bp);
  \definecolor{strokecol}{rgb}{0.0,0.0,0.0};
  \pgfsetstrokecolor{strokecol}
  \draw (6.0422bp,108.9bp) node {};
  \pgfsetcolor{blue}
  \draw [->] (35.51bp,181.49bp) .. controls (35.688bp,171.42bp) and (35.92bp,159.1bp)  .. (36.167bp,148.06bp) .. controls (36.213bp,146.01bp) and (36.262bp,143.9bp)  .. (36.599bp,130.5bp);
  \definecolor{strokecol}{rgb}{0.0,0.0,0.0};
  \pgfsetstrokecolor{strokecol}
  \draw (61.667bp,155.93bp) node {$\frac{1}{\sqrt{2}}$};
  \pgfsetcolor{red}
  \draw [->,dotted] (31.103bp,88.243bp) .. controls (28.94bp,77.584bp) and (28.169bp,64.486bp)  .. (33.417bp,54.0bp) .. controls (37.934bp,44.976bp) and (46.005bp,37.678bp)  .. (63.983bp,26.228bp);
  \definecolor{strokecol}{rgb}{0.0,0.0,0.0};
  \pgfsetstrokecolor{strokecol}
  \draw (36.042bp,61.875bp) node {};
  \pgfsetcolor{blue}
  \draw [->] (46.077bp,89.484bp) .. controls (52.311bp,76.595bp) and (60.724bp,59.199bp)  .. (72.562bp,34.723bp);
  \definecolor{strokecol}{rgb}{0.0,0.0,0.0};
  \pgfsetstrokecolor{strokecol}
  \draw (64.667bp,61.875bp) node {$\frac{i}{\sqrt{2}}$};
  \pgfsetcolor{red}
  \draw [->,dotted] (108.72bp,181.53bp) .. controls (108.47bp,169.85bp) and (108.15bp,154.95bp)  .. (107.61bp,130.28bp);
  \definecolor{strokecol}{rgb}{0.0,0.0,0.0};
  \pgfsetstrokecolor{strokecol}
  \draw (110.04bp,155.93bp) node {};
  \pgfsetcolor{blue}
  \draw [->] (121.32bp,185.51bp) .. controls (131.61bp,171.58bp) and (146.51bp,151.41bp)  .. (164.89bp,126.53bp);
  \definecolor{strokecol}{rgb}{0.0,0.0,0.0};
  \pgfsetstrokecolor{strokecol}
  \draw (152.79bp,155.93bp) node {-1};
  \pgfsetcolor{red}
  \draw [->,dotted] (96.007bp,90.666bp) .. controls (92.39bp,84.323bp) and (88.72bp,76.93bp)  .. (86.417bp,69.75bp) .. controls (84.167bp,62.734bp) and (82.719bp,54.932bp)  .. (80.69bp,36.26bp);
  \definecolor{strokecol}{rgb}{0.0,0.0,0.0};
  \pgfsetstrokecolor{strokecol}
  \draw (88.042bp,61.875bp) node {};
  \pgfsetcolor{blue}
  \draw [->] (101.18bp,88.174bp) .. controls (97.437bp,75.864bp) and (92.572bp,59.846bp)  .. (85.236bp,35.69bp);
  \definecolor{strokecol}{rgb}{0.0,0.0,0.0};
  \pgfsetstrokecolor{strokecol}
  \draw (118.79bp,61.875bp) node {$\frac{1}{\sqrt{2}}$};
  \pgfsetcolor{red}
  \draw [->,dotted] (169.49bp,88.929bp) .. controls (164.22bp,77.712bp) and (156.34bp,63.827bp)  .. (146.17bp,54.0bp) .. controls (135.27bp,43.475bp) and (120.58bp,35.268bp)  .. (97.261bp,24.966bp);
  \definecolor{strokecol}{rgb}{0.0,0.0,0.0};
  \pgfsetstrokecolor{strokecol}
  \draw (182.67bp,61.875bp) node {$\frac{1}{\sqrt{2}}$};
  \pgfsetcolor{blue}
  \draw [->] (194.14bp,95.755bp) .. controls (207.66bp,84.577bp) and (222.71bp,67.707bp)  .. (212.17bp,54.0bp) .. controls (199.7bp,37.793bp) and (144.1bp,27.536bp)  .. (98.19bp,21.086bp);
  \definecolor{strokecol}{rgb}{0.0,0.0,0.0};
  \pgfsetstrokecolor{strokecol}
  \draw (217.04bp,61.875bp) node {};
  \pgfsetcolor{red}
  \draw [->] (71.167bp,370.21bp) .. controls (71.167bp,358.85bp) and (71.167bp,343.46bp)  .. (71.167bp,318.42bp);
  \definecolor{strokecol}{rgb}{0.0,0.0,0.0};
  \pgfsetstrokecolor{strokecol}
  \draw (73.042bp,344.05bp) node {};
\begin{scope}
  \definecolor{strokecol}{rgb}{1.0,0.0,0.0};
  \pgfsetstrokecolor{strokecol}
  \draw (71.17bp,297.02bp) ellipse (21.15bp and 21.15bp);
  \definecolor{strokecol}{rgb}{0.0,0.0,0.0};
  \pgfsetstrokecolor{strokecol}
  \draw (71.167bp,297.02bp) node {$y_2$};
\end{scope}
\begin{scope}
  \definecolor{strokecol}{rgb}{1.0,0.0,0.0};
  \pgfsetstrokecolor{strokecol}
  \draw (35.17bp,202.96bp) ellipse (21.15bp and 21.15bp);
  \definecolor{strokecol}{rgb}{0.0,0.0,0.0};
  \pgfsetstrokecolor{strokecol}
  \draw (35.167bp,202.96bp) node {$y_1$};
\end{scope}
\begin{scope}
  \definecolor{strokecol}{rgb}{1.0,0.0,0.0};
  \pgfsetstrokecolor{strokecol}
  \draw (109.17bp,202.96bp) ellipse (21.15bp and 21.15bp);
  \definecolor{strokecol}{rgb}{0.0,0.0,0.0};
  \pgfsetstrokecolor{strokecol}
  \draw (109.17bp,202.96bp) node {$y_1$};
\end{scope}
\begin{scope}
  \definecolor{strokecol}{rgb}{1.0,0.0,0.0};
  \pgfsetstrokecolor{strokecol}
  \draw (80.17bp,18.0bp) ellipse (18.0bp and 18.0bp);
  \definecolor{strokecol}{rgb}{0.0,0.0,0.0};
  \pgfsetstrokecolor{strokecol}
  \draw (80.167bp,18.0bp) node {1};
\end{scope}
\begin{scope}
  \definecolor{strokecol}{rgb}{1.0,0.0,0.0};
  \pgfsetstrokecolor{strokecol}
  \draw (37.17bp,108.9bp) ellipse (21.15bp and 21.15bp);
  \definecolor{strokecol}{rgb}{0.0,0.0,0.0};
  \pgfsetstrokecolor{strokecol}
  \draw (37.167bp,108.9bp) node {$y_0$};
\end{scope}
\begin{scope}
  \definecolor{strokecol}{rgb}{1.0,0.0,0.0};
  \pgfsetstrokecolor{strokecol}
  \draw (107.17bp,108.9bp) ellipse (21.15bp and 21.15bp);
  \definecolor{strokecol}{rgb}{0.0,0.0,0.0};
  \pgfsetstrokecolor{strokecol}
  \draw (107.17bp,108.9bp) node {$y_0$};
\end{scope}
\begin{scope}
  \definecolor{strokecol}{rgb}{1.0,0.0,0.0};
  \pgfsetstrokecolor{strokecol}
  \draw (177.17bp,108.9bp) ellipse (21.15bp and 21.15bp);
  \definecolor{strokecol}{rgb}{0.0,0.0,0.0};
  \pgfsetstrokecolor{strokecol}
  \draw (177.17bp,108.9bp) node {$y_0$};
\end{scope}
\end{tikzpicture}}

    \caption{An example of FBDD \cite{tanaka_quantum_2024} representing the quantum state     $\frac{2}{\sqrt{23}}[1,1,\frac{1}{\sqrt{2}},\frac{i}{2},-1,-\frac{1}{\sqrt{2}},\frac{1}{\sqrt{2}},1]^T$. The coefficient $\frac{2}{\sqrt{23}}$ was omitted here.
    }
    \label{fig:fbdd}
\end{figure}

In this section, we address the problem of quantum state preparation with access to a sufficient number of ancilla qubits. Our approach is inspired by the work of Tanaka et al. \cite{tanaka_quantum_2024}, which introduced an efficient algorithm for quantum state preparation (QSP) using decision diagrams. Their algorithm begins with the \(\ket{0}\) state and prepares the target quantum state by traversing the decision diagram in a breadth-first manner. For each non-terminal node, an ancilla qubit is employed to mark the node as open or closed, and the algorithm processes the node accordingly. Consequently, both the time complexity and the gate complexity of the resulting quantum circuit scale with the number of nodes in the decision diagram, and $\mathrm{size}(\F)$ ancilla qubits are required, where $\mathrm{size}(\F)$ denotes the number of non-terminal nodes in the decision diagram $\F$.

However, the decision diagram used in \cite{tanaka_quantum_2024} is a weighted free binary decision diagram (WFBDD), which is also less compact compared to the LimTDD representation. For instance, the WFBDD shown in Fig.~\ref{fig:fbdd} has 7 nodes, whereas the LimTDD representation has only 6 nodes. To bridge this gap, we have designed a LimTDD-based QSP algorithm that also utilises $\mathrm{size}(\F)$ ancilla qubits.

\subsection{Algorithm}

Our algorithm, described in Alg.~\ref{alg:State_Pre_mula}, begins by eliminating the operator on the incoming edge of the root node. For any non-terminal node \( v \), an ancilla qubit \( q_{a_v} \) is assigned to it. Initially, the ancilla qubit corresponding to the root node is set to \( \ket{1} \), marking the root node as open, while the ancilla qubits corresponding to other non-terminal nodes are set to \( \ket{0} \).

Consider a non-terminal node \( v \) with the form
\[
\Circled{v_0} \overset{I}{\dashleftarrow} \Circled{v} \xrightarrow{\lambda O} \Circled{v_1}.
\]
Whenever a predecessor node \( v' \) marked open by \( q_{a_{v'}} \) is reached, and \( v \) is the \( b \)-successor of \( v' \), where $b\in \{0, 1\}$, \( v \) can be marked open with the control condition \( \ket{1}_{a_{v'}} \ket{b}_{v'} \) along the path. After all predecessors have been processed, the node \( v \) will be marked open by \( q_{a_v} \), and any of its brother nodes will be marked closed. This can be represented as
\[
\ket{1}_{a_v} \ket{p} \ket{v} + \ket{0}_{a_v} \ket{\rm Res},
\]
where \( \ket{\rm Res} \) is orthogonal to \( \ket{p} \ket{*} \), and \( \ket{p} \) represents all possible prefix paths leading to \( v \).

We first address the operator \( O \) on the high-edge of \( v \). This can be eliminated using a Controlled-Controlled-\( O^\dag \) gate with the control condition \( \ket{1}_{a_v} \ket{1}_v \), transforming the state to
\[
\ket{1}_{a_v} \ket{p} \big(\ket{0}_v \ket{v_0} + \lambda \ket{1}_v \ket{v_1}\big) + \ket{0}_{a_v} \ket{\rm Res}.
\]
At this point, the ancilla qubits \( q_{a_{v_0}} \) and \( q_{a_{v_1}} \) for \( v_0 \) and \( v_1 \) are in the state \( \ket{0} \) along the path \( p \). We then apply two \( CCX \) gates with control conditions \( \ket{1}_{a_v} \ket{0}_v \) and \( \ket{1}_{a_v} \ket{1}_v \) and target qubits \( q_{a_{v_0}} \) and \( q_{a_{v_1}} \) to mark \( v_0 \) and \( v_1 \) open along the path. The state can be simplified to
\[
\ket{1}_{a_v} \ket{p} (\ket{0}_v \ket{1}_{a_{v_0}} \ket{v_0} + \lambda \ket{1}_v \ket{1}_{a_{v_1}} \ket{v_1}) + \ket{0}_{a_v} \ket{\rm Res}.
\]
After marking all prefix paths leading to \( v_0 \) or \( v_1 \), we process \( v_0 \) and \( v_1 \) by first eliminating the operators on their high-edges, then processing their successor nodes, and finally adjusting the outgoing weights. Ultimately, both successor nodes are transformed into \( \ket{0}^{\otimes k} \) for some \( k \). The state then becomes
\[
\ket{1}_{a_v} \ket{p} \big(||v_0|| \ket{0}_v \ket{1}_{a_{v_0}} + \lambda ||v_1|| \ket{1}_v \ket{1}_{a_{v_1}}\big) \ket{0}^{\otimes k} + \ket{0}_{a_v} \ket{\rm Res'}.
\]
We then apply two additional \( CCX \) gates with control conditions \( \ket{1}_{a_v} \ket{0}_v \) and \( \ket{1}_{a_v} \ket{1}_v \) and target qubits \( q_{a_{v_0}} \) and \( q_{a_{v_1}} \) to unmark \( v_0 \) and \( v_1 \) along the path, transforming the state to
\[
\ket{1}_{a_v} \ket{p} \big(||v_0|| \ket{0}_v + \lambda ||v_1|| \ket{1}_v\big) \ket{0}^{\otimes k} + \ket{0}_{a_v} \ket{Res'}.
\]
Finally, we apply an operator
\[
\frac{1}{\sqrt{1 + |c|^2}}
\begin{bmatrix}
1 & c^\dag \\
-c & 1 \\
\end{bmatrix}
\]
with \( c = \lambda \cdot ||v_1|| / ||v_0|| \) and control condition \( \ket{1}_{a_v} \) to transform the state to
\[
||v|| \ket{1}_{a_v} \ket{p} \ket{0}^{\otimes k+1} + \ket{0}_{a_v} \ket{\rm Res'}.
\]
Note that the state \( \ket{\rm Res} \) has changed to \( \ket{\rm Res'} \), as there could be components connecting to nodes in the sub-tree of \( v \). But it remains orthogonal to \( \ket{p} \ket{*} \).

\begin{algorithm}
\caption{$\textsc{StatePre3}(v)$}
\begin{algorithmic}[1]
\Require{The root node $v$ of an LimTDD $\F$ representing an $n$-qubit quantum state $\ket{\psi}$, suppose there are $m$ non-terminal nodes in $\F$, and the qubit and ancilla qubit corresponding to $v$ are denoted as $q_v$ and $q_{a_v}$, respectively. }

\Ensure{A quantum circuit $C$, corresponding to an unitary matrix $U$, such that $U \ket{0}^{\otimes m-1}\ket{1}\ket{\psi} = \ket{0}^{\otimes m-1}\ket{1}\ket{0}^{\otimes n}$.}
\vspace{0.4em}

\State $Q \gets \{v\}$ {\color{gray}\Comment{An queue (First-In-First-Out) initialled with the node $v$}}
\State $S \gets [\ ] $ {\color{gray}\Comment{An empty stack (Last-In-First-Out)}}
\State $E \gets$ the set of all edges in $\F$ 

\State $cir \gets \mathrm{QuantumCircuit}(n+m)$


\While{$Q$ is not empty}
    \State Remove a node $v$ from $Q$
    \State Push $v$ to the stack $S$
    \State Suppose $\w\big((v,\high(v))\big)=\lambda_v \cdot O_v$
    \State Append $cir$ with a controlled $O_v^\dag$ gate with the control condition $\ket{1}_v\ket{1}_{a_v}$ 
    


    \For{$b = 0 \text{ to } 1$}
    \State $u \gets$ $b$-successor of $v$
    \State Remove edge $(v,u)$ from $E$
    \If{$u$ is the terminal node}
    \State \textbf{pass} {\color{gray}\Comment{Nothing need to be done}}
    \Else
    \If{$u$ has no incoming edge}
    \State Add $u$ to $Q$
    \EndIf
    \State Append $cir$ with a $CCX$ gate with control condition $\ket{b}_v\ket{1}_{a_v}$, and target qubit $q_{a_u}$
    \EndIf
    \EndFor
    
\EndWhile

\While{$S$ is not empty}
    \State Remove a node $v$ from $S$

    \For{$b = 0 \text{ to } 1$}
    \State $u \gets$ $b$-successor of $v$
    \If{$u$ is not the terminal node}
    \State Append $cir$ with a $CCX$ gate with control condition $\ket{b}_v\ket{1}_{a_v}$, and target qubit $q_{a_u}$
    \EndIf
    \EndFor
    \State $w_0 \gets ||\low(v)||$
    \State $w_1 \gets \lambda_v \cdot ||\high(v)||$
    \State $c \gets w_1/w_0$
    \State Append a controlled $\frac{1}{\sqrt{1+|c|^2}}\left[\begin{array}{cccc} 
		1 & c^\dag\\ 
		-c & 1\\
    \end{array}\right]$, with the control condition $\ket{1}_{a_v}$ and target $q_v$
    \vspace{0.2em}
    \State $||v|| \gets \sqrt{|w_0|^2+|w_1|^2}$.
\EndWhile
\State \Return $cir$
\end{algorithmic}
\label{alg:State_Pre_mula} 
\end{algorithm}

\subsection{An Example}\label{sec:exp}

\begin{figure*}[htbp]
\centering

\subfigure[]{
    \resizebox{0.2\textwidth}{!}{




\begin{tikzpicture}[
>=latex,
line join=bevel,
every node/.style={minimum size=1.4cm, inner sep=0pt,thick, font=\LARGE}, 
every path/.style={line width=1.5pt} 
]
\node (1) at (128.67bp,18.0bp) [draw=red,circle,font=\Huge] {$1$};
  \node (y01) at (46.668bp,108.9bp) [draw=red,circle,font=\Huge] {$v_{00}$};
  \node (y02) at (185.67bp,108.9bp) [draw=red,circle,font=\Huge] {$v_{01}$};
  \node (y10) at (66.668bp,202.96bp) [draw=red,circle,font=\Huge] {$v_{10}$};
  \node (y11) at (165.67bp,202.96bp) [draw=red,circle,font=\Huge] {$v_{11}$};
  \node (y2) at (107.67bp,297.02bp) [draw=red,circle,font=\Huge] {$v_{20}$};
  \node (-0) at (107.67bp,387.93bp) [draw,draw=none] {};
  \draw [red,->,dotted] (y01) ..controls (29.551bp,79.314bp) and (25.128bp,64.774bp)  .. (32.668bp,54.0bp) .. controls (47.583bp,32.686bp) and (76.911bp,24.306bp)  .. (1);
  \definecolor{strokecol}{rgb}{0.0,0.0,0.0};
  \pgfsetstrokecolor{strokecol}
  \draw (53.668bp,61.875bp) node {};
  \draw [blue,->] (y01) ..controls (73.917bp,78.36bp) and (94.182bp,56.39bp)  .. (1);
  \draw (116.67bp,61.875bp) node {};
  \draw [red,->,dotted] (y02) ..controls (166.06bp,77.316bp) and (153.94bp,58.425bp)  .. (1);
  \draw (180.67bp,61.875bp) node {};
  \draw [blue,->] (y02) ..controls (205.77bp,80.723bp) and (212.16bp,65.699bp)  .. (204.67bp,54.0bp) .. controls (194.36bp,37.899bp) and (174.56bp,29.102bp)  .. (1);
  \draw (230.29bp,61.875bp) node {$\frac{1}{\sqrt{2}}$};
  \draw [red,->,dotted] (y10) ..controls (31.234bp,190.19bp) and (11.932bp,180.12bp)  .. (2.6676bp,163.81bp) .. controls (-5.0407bp,150.24bp) and (6.2358bp,136.64bp)  .. (y01);
  \draw (23.668bp,155.93bp) node {};
  \draw [blue,->] (y10) ..controls (59.707bp,169.93bp) and (56.313bp,154.3bp)  .. (y02);
  \draw (105.543bp,155.93bp) node {$\frac{1}{\sqrt{2}} S$};
  \draw [red,->,dotted] (y11) ..controls (163.61bp,171.29bp) and (164.02bp,158.88bp)  .. (166.67bp,148.06bp) .. controls (167.5bp,144.65bp) and (168.68bp,141.19bp)  .. (y02);
  \draw (187.67bp,155.93bp) node {};
  \draw [blue,->] (y11) ..controls (193.01bp,184.7bp) and (203.51bp,175.28bp)  .. (208.67bp,163.81bp) .. controls (212.85bp,154.5bp) and (210.23bp,144.22bp)  .. (y02);
  \draw (230.42bp,155.93bp) node {$X$};
  \draw [red,->,dotted] (y2) ..controls (93.74bp,264.75bp) and (86.177bp,247.77bp)  .. (y10);
  \draw (110.67bp,249.99bp) node {};
  \draw [blue,->] (y2) ..controls (125.36bp,273.0bp) and (130.96bp,265.2bp)  .. (135.67bp,257.87bp) .. controls (140.87bp,249.76bp) and (146.13bp,240.69bp)  .. (y11);
  \draw (172.92bp,249.99bp) node {$Z\otimes I$};
  \draw [red,->] (-0) ..controls (107.67bp,358.85bp) and (107.67bp,343.46bp)  .. (y2);
  \draw (154.54bp,344.05bp) node {$\frac{2}{\sqrt{23}} Z\otimes I\otimes I$};
\end{tikzpicture}

%
%
}
}
\subfigure[]{
\begin{tikzpicture}
  \begin{yquant}[register/minimum height=5.2mm, operator/separation=0.9mm, control style={radius=0.8pt}, subcircuit box style={dashed}]
    qubit {$\ket{0}_{a_{01}}$} a0';
    qubit {$\ket{0}_{a_{00}}$} a0;
    qubit {$\ket{0}_{a_{11}}$} a1';
    qubit {$\ket{0}_{a_{10}}$} a1;
    qubit {$\ket{1}_{a_{20}}$} a2;
    qubit {$\ket{q_2}$} q2;
    qubit {$\ket{q_1}$} q1;
    qubit {$\ket{q_0}$} q0;
    box {$Z$} q2;
    z q1 | q2, a2;
    x a1 | a2, ~q2;
    x a1' | a2, q2;
    box {$S^\dag$} q0 | a1, q1;
    x a0 | a1, ~q1;
    x a0' | a1, q1;
    x q0 | a1', q1;
    x a0' | a1', q1;
    x a0' | a1', ~q1;
    box {$V$} q0 | a0';
    box {$U$} q0 | a0; 
    x a0' | a1', q1;
    x a0' | a1', ~q1;
    box {$U$} q1 | a1';
    x a0 | a1, ~q1;
    x a0' | a1, q1;
    box {$W$} q1 | a1;
    x a1 | a2, ~q2;
    x a1' | a2, q2;
    box {$M$} q2 | a2;
  \end{yquant}
\end{tikzpicture}}
\caption{The quantum circuit that transforms the quantum state $\ket{0}^{\otimes 5}\ket{1}\ket{\F}$ into $\ket{0}^{\otimes 5}\ket{1}\ket{000}$. In this circuit $U = \frac{1}{\sqrt{2}}\left[\begin{smallmatrix} 
		1 & 1\\ 
		-1 & 1\\
\end{smallmatrix}\right]$, $V = \frac{\sqrt{2}}{\sqrt{3}}\left[\begin{smallmatrix} 
		1 & \frac{1}{\sqrt{2}}\\ 
		-\frac{1}{\sqrt{2}} & 1\\
\end{smallmatrix}\right]$, $W = \frac{1}{\sqrt{11}}\left[\begin{smallmatrix} 
		2\sqrt{2} & \sqrt{3}\\ 
		-\sqrt{3} & 2\sqrt{2}\\
\end{smallmatrix}\right]$, $M = \frac{1}{\sqrt{23}}\left[\begin{smallmatrix} 
		\sqrt{11} & \sqrt{12}\\ 
		-\sqrt{12} & \sqrt{11}\\
\end{smallmatrix}\right]$.
}
\label{fig:pre_cir_mula}
\end{figure*}

The circuit for preparing the quantum state corresponding to the quantum state shown in Fig. \ref{fig:limtdd} was given in Fig. \ref{fig:pre_cir_mula}. In the following part, we introduce the procedure step-by-step. The initial state is in $\frac{2}{\sqrt{23}}\ket{0}^{\otimes 4}\ket{1}(Z\otimes I \otimes I \ket{v_{20}})$. We will omit the state of the ancilla qubits if it is in $\ket{0}$ for convenience. Also, we will omit the state $\ket{1}_{a_{20}}$, since it is always in $\ket{1}$.


    

    


\begin{enumerate}
    \item \textbf{Cancel the Operator on the Incoming Edge}:
    \begin{itemize}
        \item Apply a $Z$ gate on qubit $ q_2 $ to cancel the $ Z \otimes I \otimes I $ operator on the incoming edge of the LimTDD. The state becomes:
        $\ket{v_{20}} $.
    \end{itemize}

    \item \textbf{Ordered traversal to cope with the Operators}:
    
    \begin{enumerate}
        \item \textbf{Process the $ v_{20} $ Node}:
        \begin{itemize}
            \item Apply a $CCZ$ gate with $ q_{a_{20}}$ and $ q_2 $ as controls and $ q_1 $ as the target to cancel the $ Z \otimes I $ operator on the high-edge of the $ v_{20} $ node. The state becomes:
            $ \ket{0} \ket{v_{10}} + \ket{1} \ket{v_{11}}$.
            \item Apply two $CCX$ gates to mark the two nodes $v_{10}$ and $v_{11}$ as open with $q_{a_{10}}$ and $q_{a_{11}}$. The state becomes:
            $ \ket{0} \ket{1}_{a_{10}}\ket{v_{10}} + \ket{1} \ket{1}_{a_{11}}\ket{v_{11}}$.
        \end{itemize}

        \item \textbf{Process the $ v_{10} $ Node}:
        \begin{itemize}
            \item Apply a CC-$S^\dag$ gate to cancel the $S$ operator on the high-edge of the $ v_{10} $ node. The state becomes:
            $ \ket{0}\ket{1}_{a_{10}} (\ket{0}\ket{v_{00}}+\frac{1}{\sqrt{2}}\ket{1}\ket{v_{11}}) + \ket{1}\ket{1}_{a_{11}} \ket{v_{11}}$.
            \item Apply two $CCX$ gates to mark the two nodes $v_{00}$ and $v_{01}$ along the prefix path 00 and 01 as open with $q_{a_{00}}$ and $q_{a_{01}}$. The state becomes:
            $ \ket{0}\ket{1}_{a_{10}} (\ket{0}\ket{1}_{a_{00}}\ket{v_{00}}+\frac{1}{\sqrt{2}}\ket{1}\ket{1}_{a_{01}}\ket{v_{01}}) + \ket{1}\ket{1}_{a_{11}} \ket{v_{11}}$.
        \end{itemize}

        \item \textbf{Process the $ v_{11} $ Node}:
        \begin{itemize}
            \item Apply a $CCX$ gate to cancel the $X$ operator on the high-edge of the $ v_{11} $ node. The state becomes:
            $ \ket{0}\ket{1}_{a_{10}} (\ket{0}\ket{1}_{a_{00}}\ket{v_{00}}+\frac{1}{\sqrt{2}}\ket{1}\ket{1}_{a_{01}}\ket{v_{01}}) + \ket{1}\ket{1}_{a_{11}} (\ket{0}+\ket{1})\ket{v_{01}}$.
            \item Apply two $CCX$ gates to mark the node $v_{01}$ along the prefix path 10 and 11 as open. The state becomes:
            $ \ket{0}\ket{1}_{a_{10}} (\ket{0}\ket{1}_{a_{00}}\ket{v_{00}}+\frac{1}{\sqrt{2}}\ket{1}\ket{1}_{a_{01}}\ket{v_{01}}) + \ket{1}\ket{1}_{a_{11}} (\ket{0}+\ket{1})\ket{1}_{a_{01}}\ket{v_{01}}$.
        \end{itemize}
        
    \end{enumerate}

    \item \textbf{Reverse order traversal to cope with the Weights}:
    
    \begin{enumerate}
        \item \textbf{Process the $ v_{01}$ and $v_{00}$ Node}:
            \begin{itemize}
            \item Apply $CV$ gate to change the state $\ket{v_{01}}$ to $\frac{\sqrt{3}}{\sqrt{2}}\ket{0}$, and apply $CU$ gate to change the state $\ket{v_{00}}$ to $\sqrt{2}\ket{0}$, the state becomes:
            $ \ket{0}\ket{1}_{a_{10}} (\sqrt{2} \ket{0}\ket{1}_{a_{00}}+\frac{\sqrt{3}}{2}\ket{1}\ket{1}_{a_{01}})\ket{0} + \frac{\sqrt{3}}{\sqrt{2}}\ket{1}\ket{1}_{a_{11}} (\ket{0}+\ket{1})\ket{1}_{a_{01}}\ket{0}$.
        \end{itemize}
        
        \item \textbf{Process the $ v_{11} $ Node}:
        \begin{itemize}
            \item Apply two $CCX$ gates to unmark the $v_{01}$ along prefix path 10 and 11, and change the state to: 
            $ \ket{0}\ket{1}_{a_{10}} (\sqrt{2} \ket{0}\ket{1}_{a_{00}}+\frac{\sqrt{3}}{2}\ket{1}\ket{1}_{a_{01}})\ket{0} + \frac{\sqrt{3}}{\sqrt{2}}\ket{1}\ket{1}_{a_{11}} (\ket{0}+\ket{1})\ket{0}$.
            \item Apply a $CU$ gate reduce the weight on the outgoing edges of $v_{11}$ and change the state to:  
            $ \ket{0}\ket{1}_{a_{10}} (\sqrt{2} \ket{0}\ket{1}_{a_{00}}+\frac{\sqrt{3}}{2}\ket{1}\ket{1}_{a_{01}})\ket{0} + \sqrt{3}\ket{1}\ket{1}_{a_{11}}\ket{0}\ket{0}$.
        \end{itemize}
        
        \item \textbf{Process the $ v_{10} $ Node}:
        \begin{itemize}
            \item Apply two $CCX$ gates to unmark $v_{00}$ and $v_{01}$ along prefix paths 00 and 01, and the state becomes:
            $ \ket{0}\ket{1}_{a_{10}} (\sqrt{2} \ket{0}+\frac{\sqrt{3}}{2}\ket{1})\ket{0} + \sqrt{3}\ket{1}\ket{1}_{a_{11}}\ket{0}\ket{0}$.
            \item Apply a $CW$ gate to reduce the weight on the outgoing edges of the $v_{10}$ node, the state becomes:
            $ \frac{\sqrt{11}}{2}\ket{0}\ket{1}_{a_{10}} \ket{0}\ket{0} + \sqrt{3}\ket{1}\ket{1}_{a_{11}}\ket{0}\ket{0}$.
        \end{itemize}
        
        \item \textbf{Process the $ v_{20} $ Node}:
        \begin{itemize}
            \item Apply two $CCX$ gates to unmark $v_{10}$ and $v_{11}$, and the state becomes:
            $ \frac{\sqrt{11}}{2}\ket{0}
            \ket{0}\ket{0} + \sqrt{3}\ket{1}
            \ket{0}\ket{0}$.
            
            \item Apply a $CM$ gate to reduce the weight on the outgoing edges of the $v_{20}$ node, the state becomes:
            $ \frac{\sqrt{23}}{2}\ket{0}\ket{0}\ket{0}$.
        \end{itemize}
        
    \end{enumerate}
\end{enumerate}

\subsection{Complexity}
\subsubsection{Time Complexity}

In this algorithm, we traverse the decision graph in a breadth-first manner, ensuring that each non-terminal node is visited exactly twice. Given that there are \( m \) non-terminal nodes, the time complexity of this algorithm is \( \mathcal{O}(m) \).

\subsubsection{Gate Complexity}

The gate complexity is determined by the operations required to eliminate high-edge operators and adjust outgoing weights. Specifically, high-edge operators are eliminated using 3-qubit controlled gates, outgoing weights are adjusted using 2-qubit controlled gates, and the operator on the incoming edge is handled using no more than \( n \) single-qubit gates. For each non-terminal node, 4 \( CCX \) gates are required to open or close its two successors.

In summary, the upper bound on the gate complexity is:
\begin{itemize}
    \item \( (3n + 4)m \) 3-qubit gates,
    \item \( m \) 2-qubit gates,
    \item \( n \) single-qubit gates.
\end{itemize}

\subsubsection{Special Case}

For decision diagrams in the tower form, the upper bound reduces to \( n \) single-qubit gates, \( n \) two-qubit gates, and \( \frac{n(n-1)}{2} + 2n \) three-qubit gates.

\subsection{Further Optimisation}

It is important to note that although an ancilla qubit is assigned to every non-terminal node, only branch nodes require ancilla qubits to maintain the described complexity in practice. The trade-off is that when processing non-branch nodes, each gate must incorporate an additional control qubit. For instance, consider a branch node \( v \) with a 0-successor \( v_0 \) that is not a branch node. When processing node \( v \), we can use the control condition \( \ket{1}_{a_v} \). However, when processing node \( v_0 \), we need to use the control condition \( \ket{1}_{a_v} \ket{0}_v \).

Furthermore, the minimum number of ancilla qubits required to maintain this complexity can be reduced to \( \lceil \log(m) \rceil + 1 \), where \( m \) is the number of branch nodes. In this scenario, each branch node is encoded with a \( \lceil \log(m) \rceil \)-qubit state. For example, suppose a node \( v \) is encoded with \( \ket{001} \), and its 0-successor and 1-successor are encoded with \( \ket{010} \) and \( \ket{011} \), respectively. Also, suppose \( v \) has been marked open as \( \ket{001} \ket{v} = \ket{001} (\ket{0}_v \ket{v_0} + \lambda O \ket{1}_v \ket{v_1}) \). Then, we can introduce another ancilla qubit \( q_a \) and adjust the entire state to \( \ket{1}_a \ket{001} (\ket{0}_v \ket{v_0} + \lambda O \ket{1}_v \ket{v_1}) + \ket{0}_a \ket{\rm Res} \). Subsequently, we can use the control conditions \( \ket{1}_a \ket{0}_v \) and \( \ket{1}_a \ket{1}_v \) to adjust the quantum state to \( \ket{1}_a (\ket{0}_v \ket{010} \ket{v_0} + \lambda O \ket{1}_v \ket{011} \ket{v_1}) + \ket{0}_a \ket{\rm Res} \). Finally, we recover \( q_a \) using the control conditions \( \ket{010} \) and \( \ket{011} \), resulting in the state \( \ket{0}_a (\ket{0}_v \ket{010} \ket{v_0} + \lambda O \ket{1}_v \ket{011} \ket{v_1}) + \ket{0}_a \ket{\rm Res} \). We then use \( \ket{010} \) and \( \ket{011} \) as control conditions to process \( v_0 \) and \( v_1 \), respectively. However, the cost is that each gate, which originally required one control qubit (\( q_{a_v} \)), now requires \( \lceil \log(m) \rceil \) control qubits. This strategy is also applicable to Alg.~\ref{alg:State_Pre_lima} introduced below.

\section{LimTDD based QSP with Optional Number of Ancilla qubits} \label{sec:Algorithms_lima}

In this section, we introduce an algorithm that can flexibly utilise up to \( m \) ancilla qubits. This algorithm serves as a bridge between Alg.~\ref{alg:State_Pre_onea} and Alg.~\ref{alg:State_Pre_mula}, leveraging the available ancilla resources as much as possible to reduce computational complexity.

\subsection{Algorithm}

This algorithm combines the procedure of Alg. ~\ref{alg:State_Pre_onea} and Alg.~\ref{alg:State_Pre_mula}. We begin by reserving one ancilla qubit for the procedure outlined in Alg. ~\ref{alg:State_Pre_onea}. If additional ancilla qubits are available, we proceed with the steps described in Alg.~\ref{alg:State_Pre_mula}; otherwise, we revert to the procedure in Algorithm~\ref{alg:State_Pre_onea}. Essentially, when the number of available ancilla qubits exceeds $\mathrm{size}(\F)$ , the algorithm defaults to Alg.~\ref{alg:State_Pre_mula}. Conversely, when only one ancilla qubit is available, the algorithm reduces to Alg.~\ref{alg:State_Pre_onea}.

Consider a non-terminal node \( v \) with the form
\[
\Circled{v_0} \overset{I}{\dashleftarrow} \Circled{v} \xrightarrow{\lambda O} \Circled{v_1},
\]
which has been marked open by \( q_{a_v} \), represented as \( \ket{1}_{a_v} \ket{v} + \ket{0}_{a_v} \ket{\rm Res} \). The operator \( O \) can be eliminated using the control condition \( \ket{1}_{a_v} \). Suppose neither \( v_0 \) nor \( v_1 \) has been assigned an ancilla qubit. The circuit generated by using Alg.~\ref{alg:State_Pre_onea} can transform \( \ket{1}_a \ket{v_0} \) and \( \ket{1}_a \ket{v_1} \) into \( ||v_0|| \ket{1}_a \ket{0}^{\otimes k} \) and \( ||v_1|| \ket{1}_a \ket{0}^{\otimes k} \), respectively. Applying these circuits with control conditions \( \ket{1}_{a_v} \ket{0}_v \) and \( \ket{1}_{a_v} \ket{1}_v \) changes the original state to
\[
\ket{1}_{a_v} \big(||v_0|| \ket{0}_v + \lambda ||v_1|| \ket{1}_v\big) \ket{0}^{\otimes k} + \ket{0}_{a_v} \ket{\rm Res}.
\]
Subsequently, applying an operator
\[
\frac{1}{\sqrt{1 + |c|^2}}
\begin{bmatrix}
1 & c^\dag \\
-c & 1 \\
\end{bmatrix}
\]
with \( c = \lambda \cdot ||v_1|| / ||v_0|| \) and control condition \( \ket{1}_{a_v} \) transforms the state to
\[
||v|| \ket{1}_{a_v} \ket{0}^{\otimes k+1} + \ket{0}_{a_v} \ket{\rm Res}.
\]
If either successor node has been assigned an ancilla qubit, we follow the procedure in Algorithm~\ref{alg:State_Pre_mula} to process the node, which involves adding further gates to transform it into \( \ket{0}^{\otimes k} \). We then unmark the node and proceed to adjust the outgoing edge weights of \( v \).

\begin{algorithm}
\caption{$\textsc{StatePre4}(v,m)$}
\begin{algorithmic}[1]
\Require{The root node $v$ of a LimTDD $\F$ representing an $n$-qubit quantum state $\ket{\psi}$, suppose there are $m$ ancilla qubits available}

\Ensure{A quantum circuit $C$, corresponding to an unitary matrix $U$, such that $U \ket{0}^{\otimes m_0}\ket{1}^{\otimes m_1} \ket{\psi} = \ket{0}^{\otimes m_0}\ket{1}^{\otimes m_1} \ket{0}^{\otimes n}$, where $m_0 = \max(0,m-2)$ and $m_1 = \min(2,m)$.}
\vspace{0.4em}
\State Directly call Alg. \ref{alg:State_Pre_onea} if $m=1$
\State $q_a \gets$ a reserved ancilla qubit {\color{gray}\Comment{for calling Alg. \ref{alg:State_Pre_onea}}} 
\State $Q \gets \{v\}$ {\color{gray}\Comment{An queue initialled with the node $v$}}
\State $S \gets [\ ] $ {\color{gray}\Comment{An empty stack}}
\State $E \gets$ the set of all edges in $\F$ 

\State $cir \gets \mathrm{QuantumCircuit}(n+m)$ 
\State $avi\_anc \gets m-1$ {\color{gray}\Comment{Available ancilla qubit num}}

\While{$Q$ is not empty}
    \State Remove a node $v$ from $Q$
    \State Push $v$ to the stack $S$
    \State Allocate $v$ with an ancilla qubit $q_{a_v}$ and set $avi\_anc \gets avi\_anc-1$, if it has not been allocated one and $avi\_anc>0$
    \State Suppose $\w\big((v,\high(v))\big)=\lambda_v \cdot O_v$
    \State Append $cir$ with a controlled $O_v^\dag$ gate with the control condition $\ket{1}_v\ket{1}_{a_v}$, if $v$ has been allocated with an ancilla $a_v$ 

    \For{$b = 0 \text{ to } 1$}
    \State $u \gets$ $b$-successor of $v$
    \State Remove edge $(v,u)$ from $E$
    \State End For Loop if $u$ is the terminal node
    \State Allocate $u$ with an ancilla qubit $q_{a_u}$ and set $avi\_anc \gets avi\_anc -1 $, if it has not been allocated one and $avi\_anc>0$
    \State Add $u$ to $Q$, if $u$ has no incoming edge and has been allocated with an ancilla qubit $q_{a_u}$
    \State Append $cir$ with a $CCX$ gate with control condition $\ket{b}_v\ket{1}_{a_v}$, and target qubit $q_{a_u}$, if $u$ has been allocated with an ancilla qubit $q_{a_u}$
    
    \EndFor
    
\EndWhile

\While{$S$ is not empty}
    \State Remove a node $v$ from $S$

    \If{$\low(v)=\high(v)\neq v_T$ and $\low(v)$ has not been allocated an ancilla qubit}
    \State $cir_l \gets \textsc{StatePre2}\big(\low(v),q_{a},[]\big)$
    \State Append $cir$ with $cir_l$ with control condition $\ket{1}_{a_v}$
    \State Continue While Loop without running lines 30-39
    \EndIf

    \For{$b = 0 \text{ to } 1$}
    \State $u \gets$ $b$-successor of $v$
    \State End For Loop if $u$ is the terminal node
    \If{$u$ has not been allocated an ancilla qubit}
    \State $cir_b \gets \textsc{StatePre2}\big(u,q_{a},[]\big)$
    \State Append $cir$ with $cir_b$ with control condition $\ket{b}_v\ket{1}_{a_v}$
    \Else
    \State Append $cir$ with a $CCX$ gate with control condition $\ket{b}_v\ket{1}_{a_v}$, and target qubit $q_{a_u}$
    \EndIf
    \EndFor
    
    \State Do Lines 31-35 from Alg. \ref{alg:State_Pre_mula}
\EndWhile
\State \Return $cir$
\end{algorithmic}
\label{alg:State_Pre_lima} 
\end{algorithm}

\subsection{An Example}\label{sec:exp}

\begin{figure*}[htbp]
\centering

\subfigure[]{
    \resizebox{0.22\textwidth}{!}{




\begin{tikzpicture}[
>=latex,
line join=bevel,
every node/.style={minimum size=1.4cm, inner sep=0pt,thick, font=\LARGE}, 
every path/.style={line width=1.5pt} 
]
\node (1) at (128.67bp,18.0bp) [draw=red,circle,font=\Huge] {$1$};
  \node (y01) at (46.668bp,108.9bp) [draw=red,circle,font=\Huge] {$v_{00}$};
  \node (y02) at (185.67bp,108.9bp) [draw=red,circle,font=\Huge] {$v_{01}$};
  \node (y10) at (66.668bp,202.96bp) [draw=red,circle,font=\Huge] {$v_{10}$};
  \node (y11) at (165.67bp,202.96bp) [draw=red,circle,font=\Huge] {$v_{11}$};
  \node (y2) at (107.67bp,297.02bp) [draw=red,circle,font=\Huge] {$v_{20}$};
  \node (-0) at (107.67bp,387.93bp) [draw,draw=none] {};
  \draw [red,->,dotted] (y01) ..controls (29.551bp,79.314bp) and (25.128bp,64.774bp)  .. (32.668bp,54.0bp) .. controls (47.583bp,32.686bp) and (76.911bp,24.306bp)  .. (1);
  \definecolor{strokecol}{rgb}{0.0,0.0,0.0};
  \pgfsetstrokecolor{strokecol}
  \draw (53.668bp,61.875bp) node {};
  \draw [blue,->] (y01) ..controls (73.917bp,78.36bp) and (94.182bp,56.39bp)  .. (1);
  \draw (116.67bp,61.875bp) node {};
  \draw [red,->,dotted] (y02) ..controls (166.06bp,77.316bp) and (153.94bp,58.425bp)  .. (1);
  \draw (180.67bp,61.875bp) node {};
  \draw [blue,->] (y02) ..controls (205.77bp,80.723bp) and (212.16bp,65.699bp)  .. (204.67bp,54.0bp) .. controls (194.36bp,37.899bp) and (174.56bp,29.102bp)  .. (1);
  \draw (230.29bp,61.875bp) node {$\frac{1}{\sqrt{2}}$};
  \draw [red,->,dotted] (y10) ..controls (31.234bp,190.19bp) and (11.932bp,180.12bp)  .. (2.6676bp,163.81bp) .. controls (-5.0407bp,150.24bp) and (6.2358bp,136.64bp)  .. (y01);
  \draw (23.668bp,155.93bp) node {};
  \draw [blue,->] (y10) ..controls (59.707bp,169.93bp) and (56.313bp,154.3bp)  .. (y02);
  \draw (105.543bp,155.93bp) node {$\frac{1}{\sqrt{2}} S$};
  \draw [red,->,dotted] (y11) ..controls (163.61bp,171.29bp) and (164.02bp,158.88bp)  .. (166.67bp,148.06bp) .. controls (167.5bp,144.65bp) and (168.68bp,141.19bp)  .. (y02);
  \draw (187.67bp,155.93bp) node {};
  \draw [blue,->] (y11) ..controls (193.01bp,184.7bp) and (203.51bp,175.28bp)  .. (208.67bp,163.81bp) .. controls (212.85bp,154.5bp) and (210.23bp,144.22bp)  .. (y02);
  \draw (230.42bp,155.93bp) node {$X$};
  \draw [red,->,dotted] (y2) ..controls (93.74bp,264.75bp) and (86.177bp,247.77bp)  .. (y10);
  \draw (110.67bp,249.99bp) node {};
  \draw [blue,->] (y2) ..controls (125.36bp,273.0bp) and (130.96bp,265.2bp)  .. (135.67bp,257.87bp) .. controls (140.87bp,249.76bp) and (146.13bp,240.69bp)  .. (y11);
  \draw (172.92bp,249.99bp) node {$Z\otimes I$};
  \draw [red,->] (-0) ..controls (107.67bp,358.85bp) and (107.67bp,343.46bp)  .. (y2);
  \draw (154.54bp,344.05bp) node {$\frac{2}{\sqrt{23}} Z\otimes I\otimes I$};
\end{tikzpicture}

%
%
}
}
\subfigure[]{
\begin{tikzpicture}
  \begin{yquant}[register/minimum height=8mm, operator/separation=3mm, control style={radius=2pt}, subcircuit box style={dashed}]
    qubit {$\ket{0}_{a_{11}}$} a1';
    qubit {$\ket{0}_{a_{10}}$} a1;
    qubit {$\ket{1}_{a_{20}}$} a2;
    qubit {$\ket{q_2}$} q2;
    qubit {$\ket{q_1}$} q1;
    qubit {$\ket{q_0}$} q0;
    qubit {$\ket{1}_a$} qa;
    box {$Z$} q2;
    z q1 | q2, a2;
    x a1 | a2, ~q2;
    x a1' | a2, q2;
    box {$S^\dag$} q0 | a1, q1;
    x q0 | a1', q1;

    box {$V$} q0 | a1', qa;
    box {$U$} q1 | a1';

    box {$U$} q0 | a1, qa,~q1;
    box {$V$} q0 | a1, qa, q1;
    box {$W$} q1 | a1;
    
    x a1 | a2, ~q2;
    x a1' | a2, q2;
    box {$M$} q2 | a2;
  \end{yquant}
\end{tikzpicture}}
\caption{The quantum circuit that transforms the quantum state $\ket{0}^{\otimes 2}\ket{1}\ket{\F}$ into $\ket{0}^{\otimes 2}\ket{1}\ket{000}$. In this circuit $U = \frac{1}{\sqrt{2}}\left[\begin{smallmatrix} 
		1 & 1\\ 
		-1 & 1\\
\end{smallmatrix}\right]$, $V = \frac{\sqrt{2}}{\sqrt{3}}\left[\begin{smallmatrix} 
		1 & \frac{1}{\sqrt{2}}\\ 
		-\frac{1}{\sqrt{2}} & 1\\
\end{smallmatrix}\right]$, $W = \frac{1}{\sqrt{11}}\left[\begin{smallmatrix} 
		2\sqrt{2} & \sqrt{3}\\ 
		-\sqrt{3} & 2\sqrt{2}\\
\end{smallmatrix}\right]$, $M = \frac{1}{\sqrt{23}}\left[\begin{smallmatrix} 
		\sqrt{11} & \sqrt{12}\\ 
		-\sqrt{12} & \sqrt{11}\\
\end{smallmatrix}\right]$.
}
\label{fig:pre_cir_lima}
\end{figure*}

The circuit for preparing the quantum state corresponding to the quantum state shown in Fig. \ref{fig:limtdd} is given in Fig. \ref{fig:pre_cir_lima}.  In the remainder of this subsection, we describe the preparation procedure step-by-step. For clarity, we will also omit the state $\ket{1}_{a_{20}}$ and the state of other ancilla qubits if it is in $\ket{0}$.


    
    

\begin{enumerate}
    \item \textbf{Cancel the Operator on the Incoming Edge}:
    \begin{itemize}
        \item Apply a $Z$ gate on qubit $ q_2 $ to cancel the $ Z \otimes I \otimes I $ operator on the incoming edge of the LimTDD. The state becomes:
        $\ket{v_{20}} $.
    \end{itemize}

    \item \textbf{Ordered traversal to cope with the Operators}:
    
    \begin{enumerate}
        \item \textbf{Process the $ v_{20} $ Node}:
        \begin{itemize}
            \item Apply a CCZ gate with $ q_{a_{20}}$ and $ q_2 $ as controls and $ q_1 $ as the target to cancel the $ Z \otimes I $ operator on the high-edge of the $ v_{20} $ node. The state becomes:
            $ \ket{0} \ket{v_{10}} + \ket{1} \ket{v_{11}}$.
            \item Apply two CCX gates to mark the two nodes $v_{10}$ and $v_{11}$ as open with $q_{a_{10}}$ and $q_{a_{11}}$. The state becomes:
            $ \ket{0} \ket{1}_{a_{10}}\ket{v_{10}} + \ket{1} \ket{1}_{a_{11}}\ket{v_{11}}$.
        \end{itemize}

        \item \textbf{Process the $ v_{10} $ Node}:
        \begin{itemize}
            \item Apply a CC-$S^\dag$ gate to cancel the $S$ operator on the high-edge of the $ v_{10} $ node. The state becomes:
            $ \ket{0}\ket{1}_{a_{10}} (\ket{0}\ket{v_{00}}+\frac{1}{\sqrt{2}}\ket{1}\ket{v_{11}}) + \ket{1}\ket{1}_{a_{11}} \ket{v_{11}}$.
        \end{itemize}

        \item \textbf{Process the $ v_{11} $ Node}:
        \begin{itemize}
            \item Apply a CCX gate to cancel the $X$ operator on the high-edge of the $ v_{11} $ node. The state becomes:
            $ \ket{0}\ket{1}_{a_{10}} (\ket{0}\ket{v_{00}}+\frac{1}{\sqrt{2}}\ket{1}\ket{v_{01}}) + \ket{1}\ket{1}_{a_{11}} (\ket{0}+\ket{1})\ket{v_{01}}$.
        \end{itemize}
        
    \end{enumerate}

    \item \textbf{Reverse order traversal to cope with the Weights}:
    
    \begin{enumerate}
        
        \item \textbf{Process the $ v_{11} $ Node}:
        \begin{itemize}
            \item Use Alg. \ref{alg:State_Pre_onea} to cope with the $v_{01}$ node, a $CV$ gate will be returned which can change the state $\ket{v_{01}}$ to $\frac{\sqrt{3}}{2}\ket{0}$, add it to the circuit with control condition $\ket{1}_{a_{11}}$, and the state will becomes:
            $ \ket{0}\ket{1}_{a_{10}} (\ket{0}\ket{v_{00}}+\frac{1}{\sqrt{2}}\ket{1}\ket{v_{01}}) + \frac{\sqrt{3}}{2}\ket{1}\ket{1}_{a_{11}} (\ket{0}+\ket{1})\ket{0}$.
        
            \item Apply a $CU$ gate reduce the weight on the outgoing edges of $v_{11}$ and change the state to:  
            $ \ket{0}\ket{1}_{a_{10}} (\ket{0}\ket{v_{00}}+\frac{1}{\sqrt{2}}\ket{1}\ket{v_{01}}) + \sqrt{3}\ket{1}\ket{1}_{a_{11}} \ket{0}\ket{0}$.
            
        \end{itemize}
        
        \item \textbf{Process the $ v_{10} $ Node}:
        \begin{itemize}
            \item Use Alg. \ref{alg:State_Pre_onea} to cope with the $v_{00}$ node and $v_{01}$ node respectively, two gates $CU$ and $CV$ will be returned, which can change state $\ket{v_{00}}$ and $\ket{v_{01}}$ to $\sqrt{2}\ket{0}$ and $\frac{\sqrt{3}}{2}\ket{0}$, respectively. Adding them to the circuit with control condition $\ket{1}_{a_{10}}\ket{0}_1$ and $\ket{1}_{a_{10}}\ket{1}_1$, and the state becomes:
            $ \ket{0}\ket{1}_{a_{10}} (\sqrt{2}\ket{0}+\frac{\sqrt{3}}{2}\ket{1})\ket{0} + \sqrt{3}\ket{1}\ket{1}_{a_{11}} \ket{0}\ket{0}$.
            \item Apply a $CW$ gate to reduce the weight on the outgoing edges of the $v_{10}$ node, the state becomes:
            $ \frac{\sqrt{11}}{2}\ket{0}\ket{1}_{a_{10}} \ket{0}\ket{0} + \sqrt{3}\ket{1}\ket{1}_{a_{11}}\ket{0}\ket{0}$.
        \end{itemize}
        
        \item \textbf{Process the $ v_{20} $ Node}:
        \begin{itemize}
            \item Apply two $CCX$ gates to unmark $v_{10}$ and $v_{11}$, and the state becomes:
            $ \frac{\sqrt{11}}{2}\ket{0}
            \ket{0}\ket{0} + \sqrt{3}\ket{1}
            \ket{0}\ket{0}$.
            
            \item Apply a $CM$ gate to reduce the weight on the outgoing edges of the $v_{20}$ node, the state becomes:
            $ \frac{\sqrt{23}}{2}\ket{0}\ket{0}\ket{0}$.
        \end{itemize}
        
    \end{enumerate}
\end{enumerate}

\subsection{Complexity}

\subsubsection{Time Complexity}

Assume that \( m \) nodes have been allocated ancilla qubits, and there are \( p \) reduced paths originating from the topmost non-terminal nodes that have not been allocated ancilla qubits. Suppose these topmost non-terminal nodes correspond to qubit \( q_k \). The time complexity of this algorithm is \( \mathcal{O}(m + kp) \).

\subsubsection{Gate Complexity}

The upper bound on the gate complexity is as follows:
\begin{itemize}
    \item \( 2p \) \( k+1 \)-qubit gates,
    \item \( 3p \) \( s \)-qubit gates for \( s \in \{4, \ldots, k\} \),
    \item \( (3n + 4)m + \frac{k(k-1)}{2}p + 3p \) 3-qubit gates,
    \item \( m + kp + 3p \) 2-qubit gates,
    \item \( n \) single-qubit gates.
\end{itemize}

\subsubsection{Special Case}

For decision diagrams in the tower form, the upper bound reduces to \( n \) single-qubit gates, \( n \) two-qubit gates, and \( \frac{n(n-1)}{2} + 2m \) three-qubit gates.


\section{Experiments}\label{sec:experiments}

We conducted experiments to compare our algorithms with many existing methods, including those implemented in Qiskit, QuICT, and ADD-based, and FBDD-based methods. Specifically, Qiskit and QuICT--which to not utilise ancilla qubits--were compared with our Alg. \ref{alg:State_Pre_noa}. The ADD-based method, which uses one ancilla qubit, was compared with our Alg. \ref{alg:State_Pre_onea}. The FBDD-based algorithm, which uses as many ancilla qubits as the number of non-terminal nodes, was compared with our Alg. \ref{alg:State_Pre_mula}. In addition, we evaluated the performance of our four algorithms with different numbers of ancilla qubits.

All experiments were conducted on a Linux server equipped with a 13th Gen Intel(R) Core(TM) i5-13600KF processor and 32GB RAM. The tested quantum states were generated using Clifford + T circuits, a commonly used circuit category. For each qubit number $n$, we generated 20 random quantum states and analysed their average performance.

We measured the running time of the algorithms and the number of multi-qubit gates in the resulting circuits (similar results were obtained when calculating the number of all gates or circuit depths). It is worth noting that the circuits generated by multiple DD-based methods contain multiple-controlled gates, while circuits generated by Qiskit and QuICT only contain $CX$ gates and single-qubit gates. For fairness, we use Qiskit to transpile the circuits generated by the DD-based methods into the set of $CX$ gates and single-qubit gates so they can be compared under the same gate set. Both pre- and post-transpilation results for time and gate complexity are reported in our experiments, denoted as "nt"(not transpiled) or "t"(transpiled), respectively. Note that the ADD-based algorithm tools have implemented their own method of converting multi-control qubit gates into $CX$ and single-qubit gates. Therefore, we used their native transpilation implementation instead of Qiskit.

\subsection{Non-ancilla Algorithm (Alg. \ref{alg:State_Pre_noa}) compared with Qiskit and QuICT.}

\begin{figure*}
    \centering
    \includegraphics[scale=0.6]{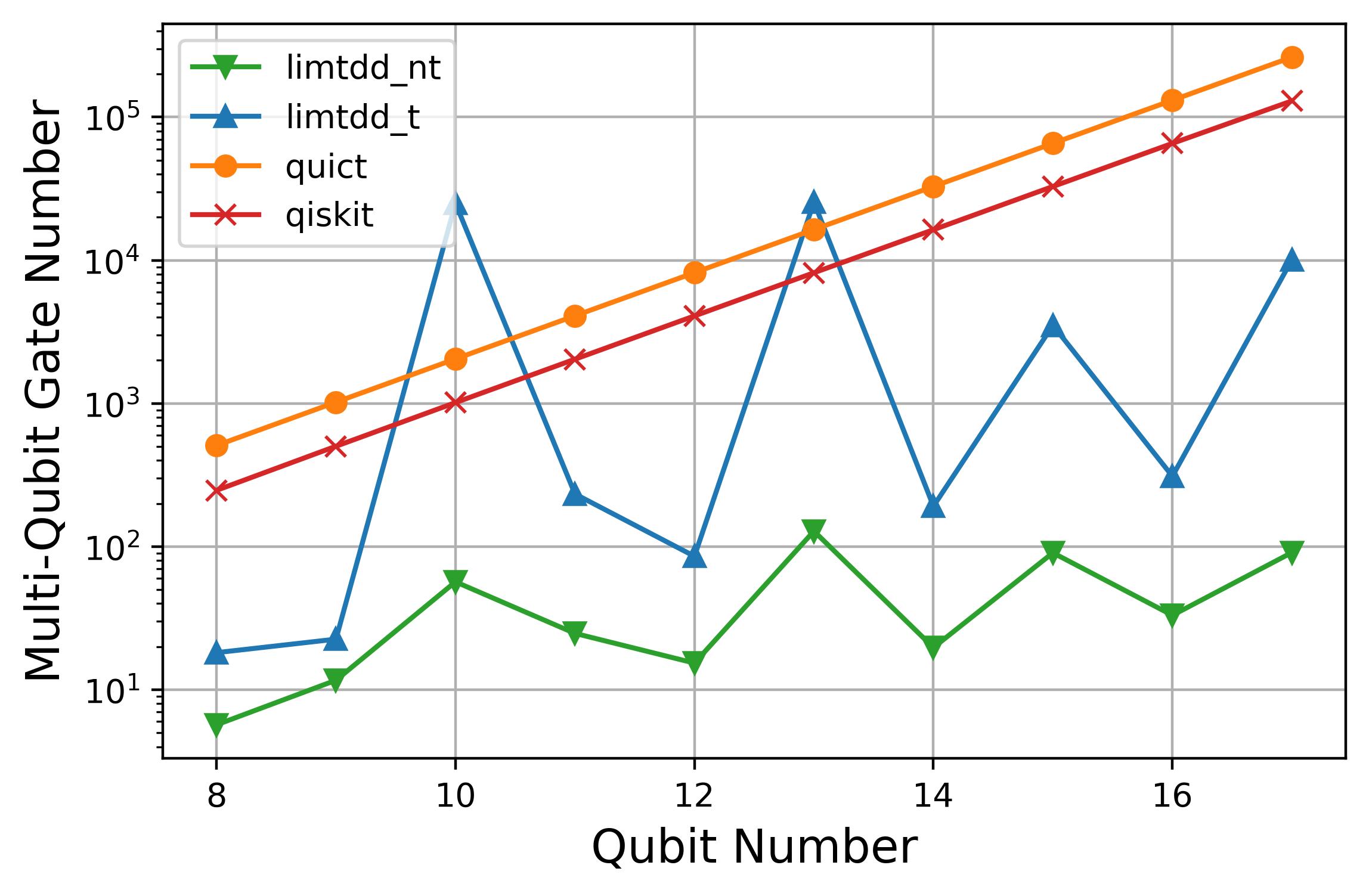}
    \includegraphics[scale=0.6]{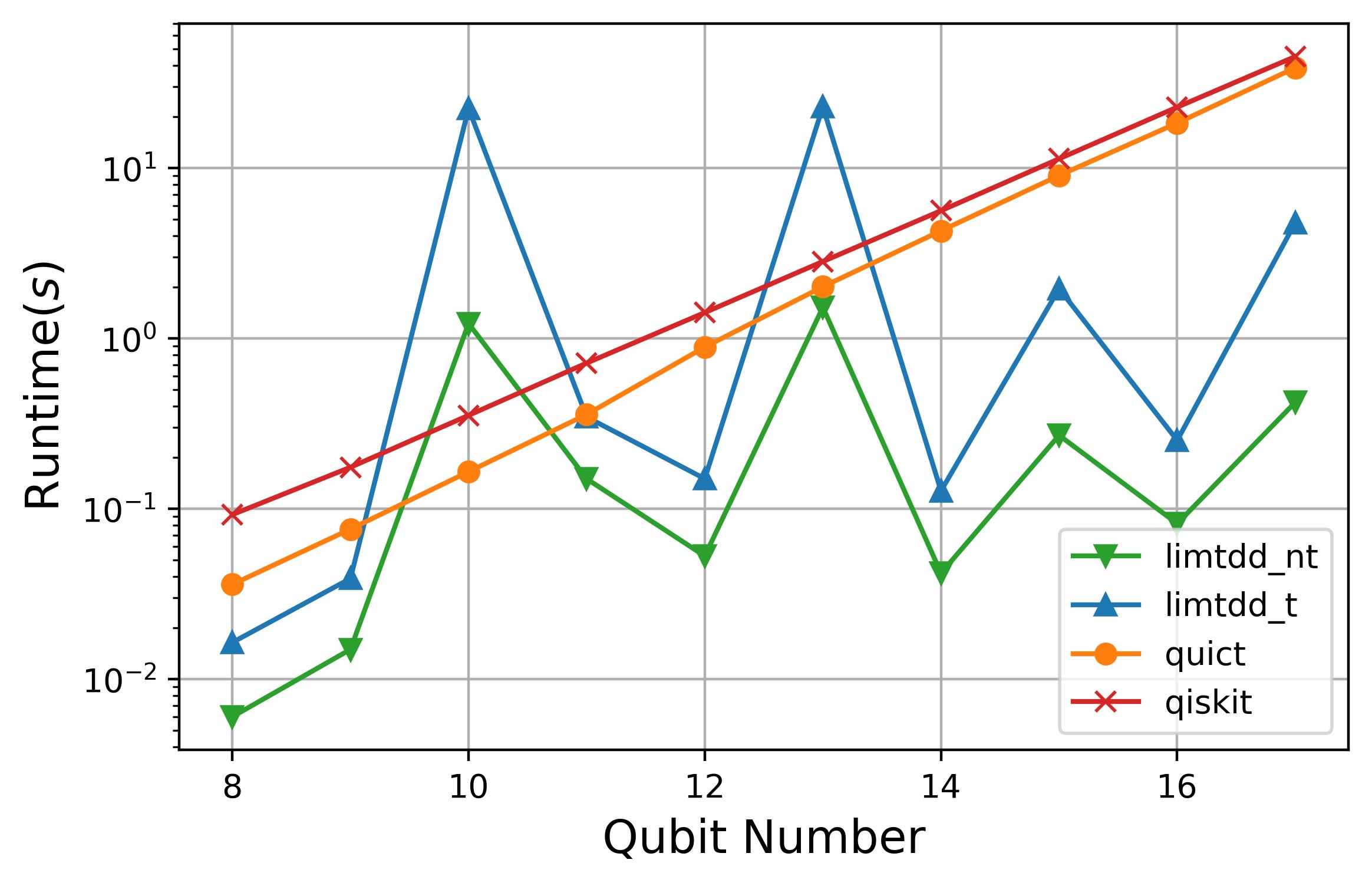}    
    \caption{Experiment results of the Alg. \ref{alg:State_Pre_noa}, compared with QuICT and Qiskit.}
    \label{fig:vs_Qis_ICT}
\end{figure*}

We first compared our no-ancilla algorithm with two widely-used quantum computing frameworks: Qiskit \cite{qiskit2024} and QuICT \cite{quict}. The algorithms used for quantum state preparation in Qiskit and QuICT are established in \cite{Iten_Colbeck_Kukuljan_Home_Christandl_2016} and \cite{mottonen2004transformation}, respectively. The experiment results are shown in Fig. \ref{fig:vs_Qis_ICT}.

\begin{itemize}
    \item \textbf{Gate Complexity}:
    Overall, our algorithm requires fewer quantum gates compared to Qiskit and QuICT, both before and after compilation. The advantage becomes more pronounced as the number of qubits increases, highlighting the scalability of our approach. For instance, for $n=15$ qubits, our method requires around 90 and 3510 gates before and after the transpilation, while both Qiskit and QuICT require around 130000 gates.

    \item \textbf{Runtime Complexity}:
    The trend of runtime complexity is consistent with that of gate complexity. Overall, when the transpilation time is excluded, our algorithm 1 requires less time. Similarly, when the number of qubits increases, the advantage becomes more pronounced and stable. For example, for $n=15$ qubits, our algorithm takes 0.27 seconds--not including an extra 1.68 seconds needed for transpilation--while Qiskit and QuICT take approximately 3 and 2 seconds, respectively. 

\end{itemize}


\subsection{One-ancilla Algorithm (Alg. \ref{alg:State_Pre_onea}) Compared with ADD-based Method }

\begin{figure*}
    \centering
    \includegraphics[scale=0.6]{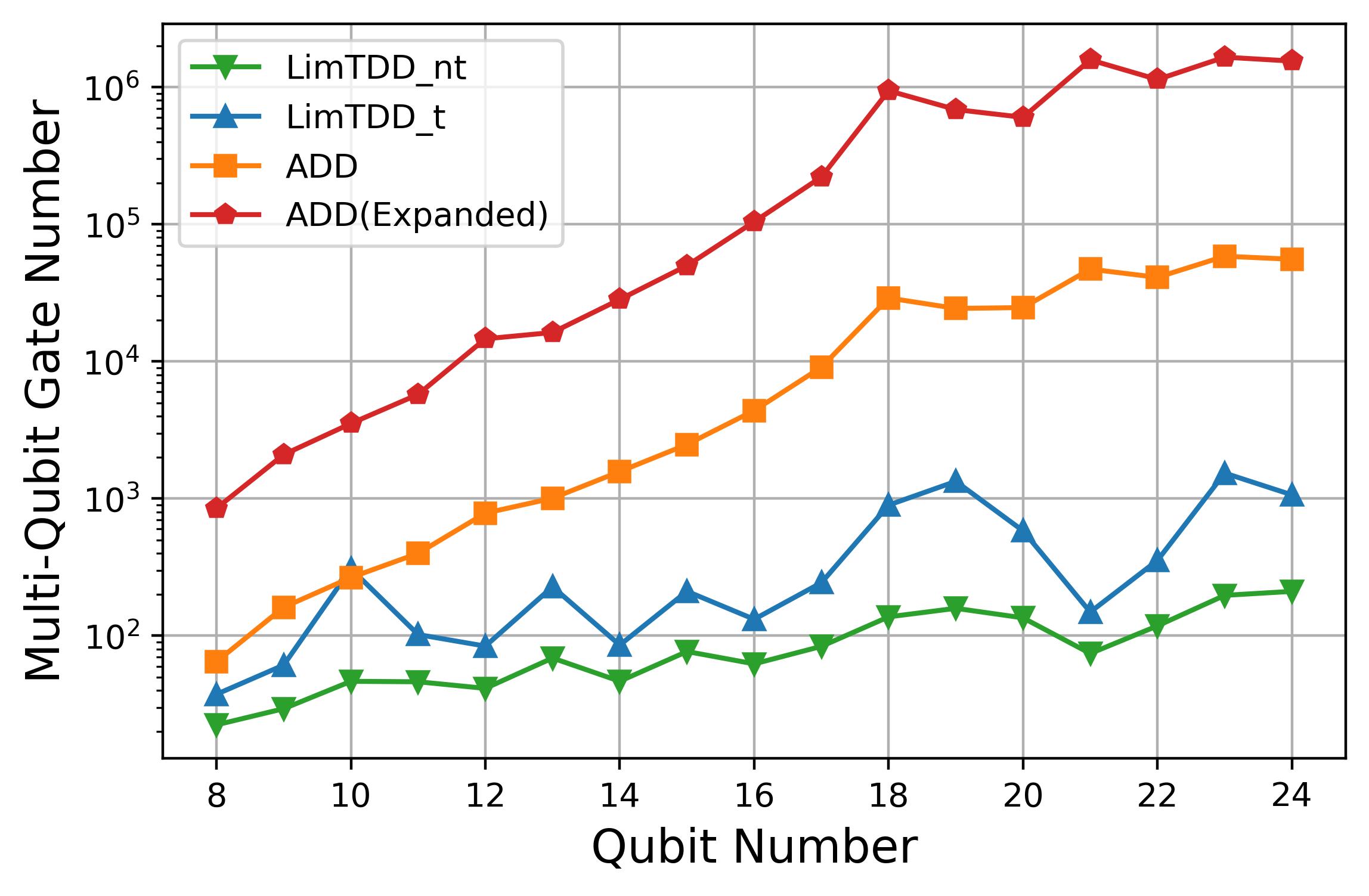}
    \includegraphics[scale=0.6]{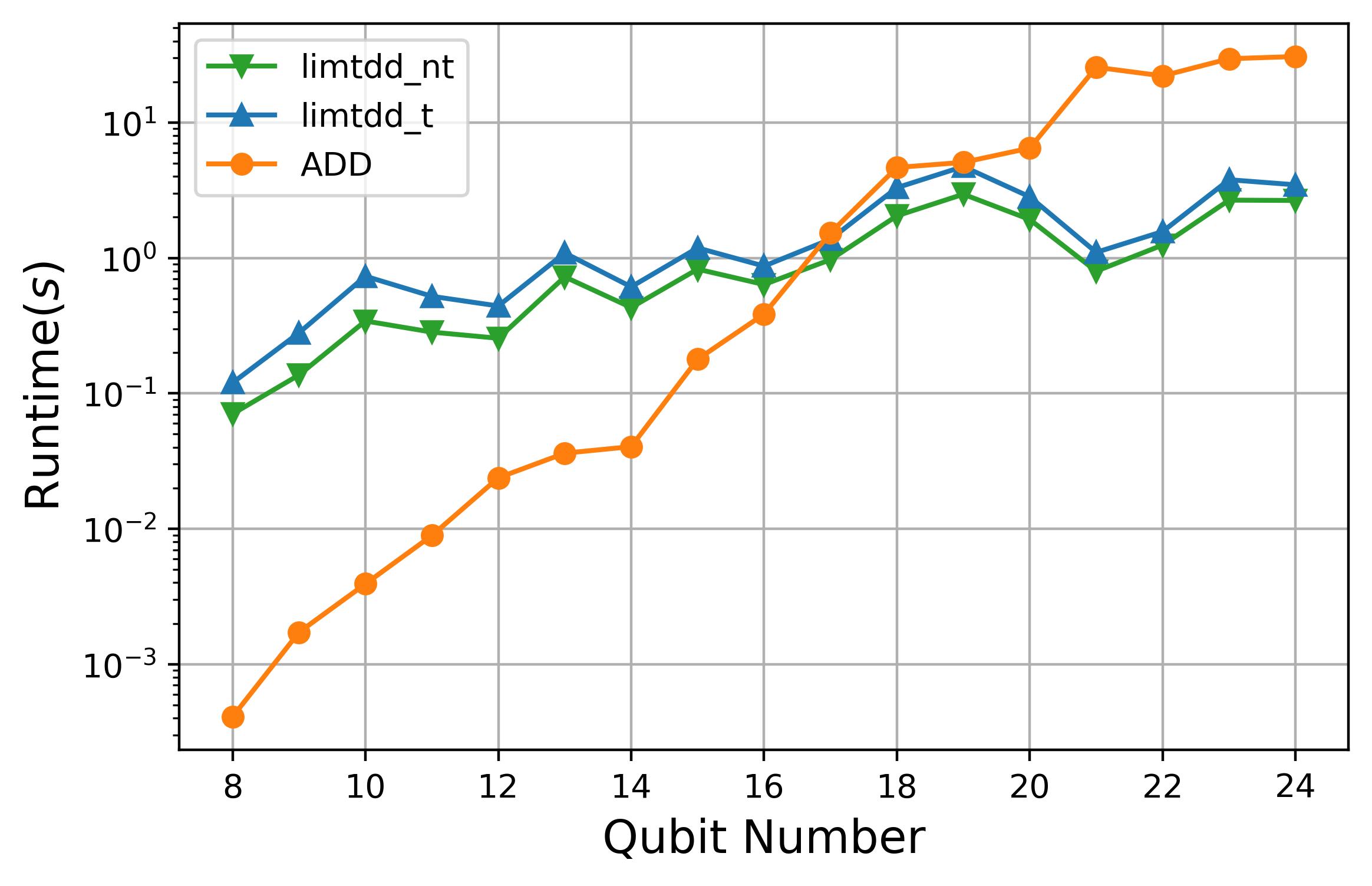}    
    \caption{Experiment results of our method against ADD-based method \cite{mozafari_efficient_2022}.}
    \label{fig:vs_ADD}
\end{figure*}

We compared our One-ancilla algorithm with the ADD-based method proposed in \cite{mozafari_efficient_2022}. Fig. \ref{fig:vs_ADD} shows the runtime and multi-qubit gate complexity of the two methods.

\begin{itemize}
    \item \textbf{Gate Complexity}: Our method consistently requires fewer gates when the number of qubits exceeds 5. This improvement is attributed to the more compact representation of quantum states using LimTDD compared to ADD. For example, for $n = 15$ qubits, our method uses approximately 80 and 200 gates before and after transpilation, while the ADD-based method uses around 2,500 and 50,000 gates before and after transpilation, respectively.
    
    \item \textbf{Runtime Complexity}: For small qubit numbers, the ADD-based method is faster due to its C++ implementation, while our Python implementation is slower until qubit number surpasses 15. As the number of qubits increases, our method's runtime becomes significantly shorter and exhibits better scalability, highlighting the advantage of LimTDD's compactness.
\end{itemize}

\subsection{Sufficient Ancilla Algorithm (Alg. \ref{alg:State_Pre_mula}) Compared with FBDD based method}


\begin{figure*}
    \centering
    \includegraphics[scale=0.6]{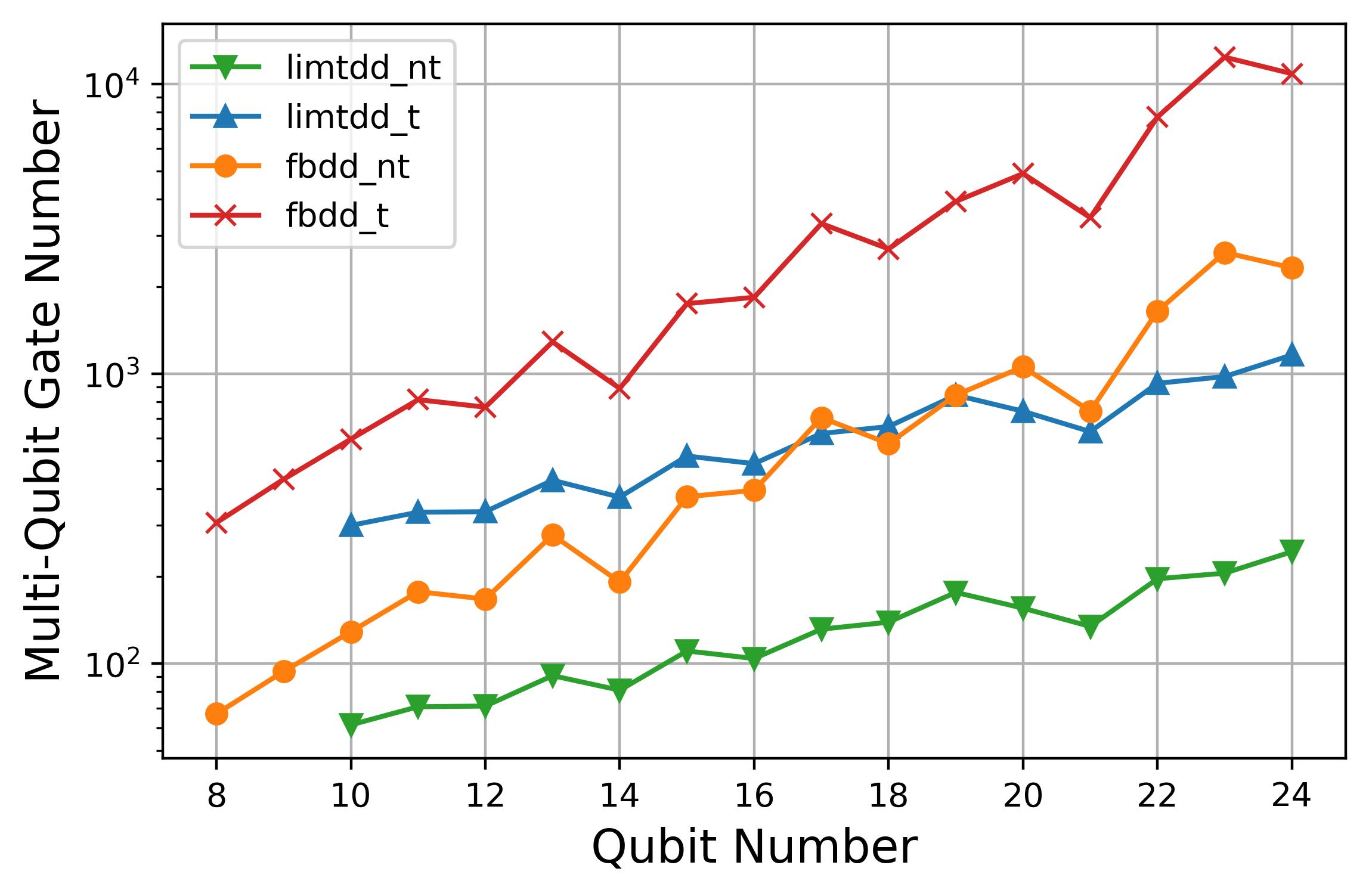}
    \includegraphics[scale=0.6]{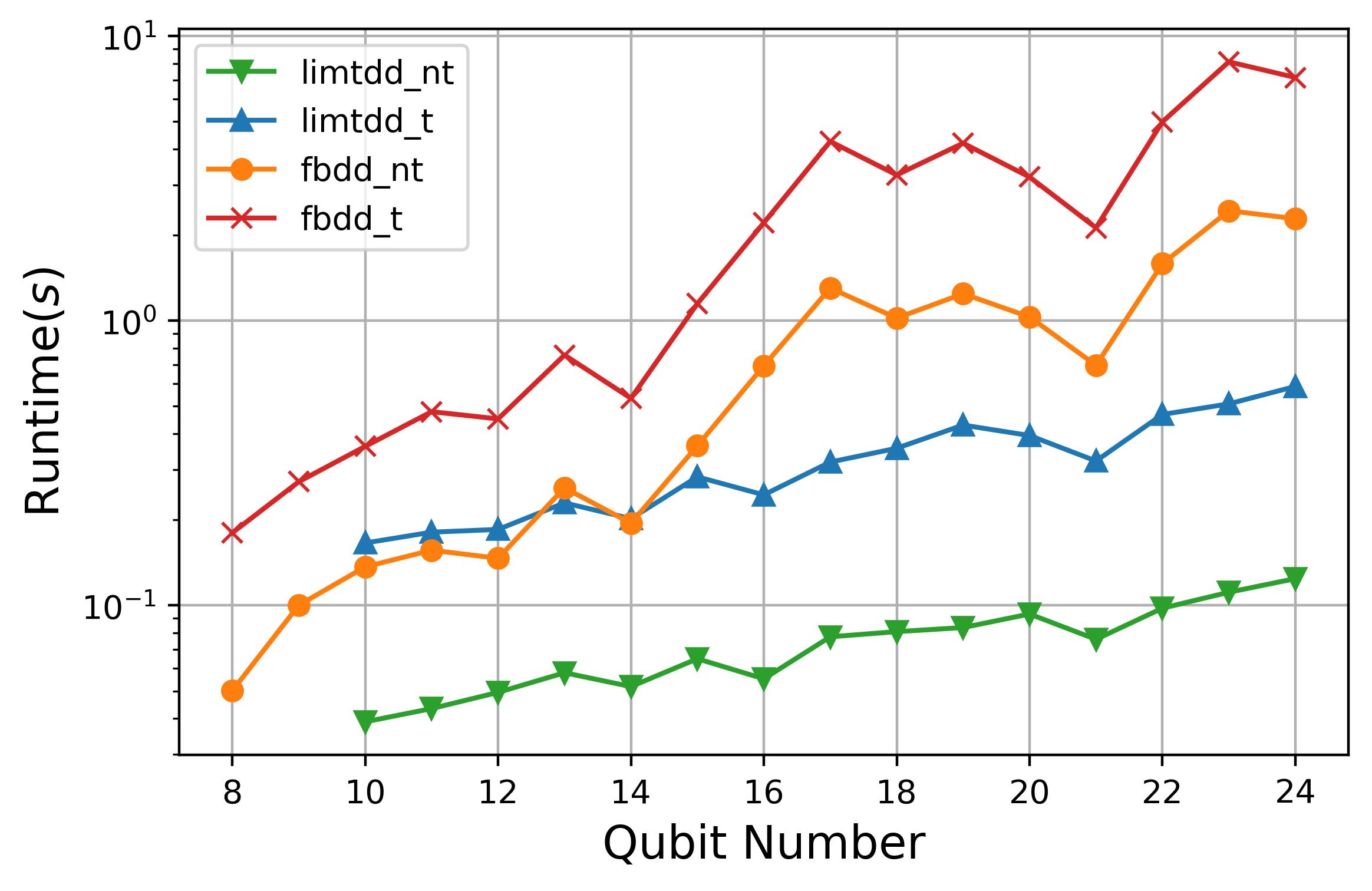}    
    \caption{Experiment results of our method (Alg. \ref{alg:State_Pre_mula}) against FBDD-based method \cite{tanaka_quantum_2024} implemented in TDD.}
    \label{fig:vs_FBDD}
\end{figure*}

We compared our sufficient-ancilla algorithm with the FBDD-based method proposed in \cite{tanaka_quantum_2024}. Note that there is no implementation provided in \cite{tanaka_quantum_2024}, so we implemented the algorithm using Tensor Decision Diagram (TDD), which can be seen as a type of FBDD. Fig. \ref{fig:vs_FBDD} shows the runtime and multi-qubit gate complexity of the two methods.

\begin{itemize}
    \item \textbf{Gate Complexity}: Our algorithm consistently requires fewer multi-qubit gates, and this trend remains after transpilation. The gap widens as the number of qubits increases.
    
    \item \textbf{Runtime Complexity}: The runtime comparison follows the same trend, further demonstrating the advantages of using LimTDD for quantum state preparation.
\end{itemize}

\subsection{The Performance of Our Four Algorithms}



\begin{figure*}
    \centering
    \includegraphics[scale=0.6]{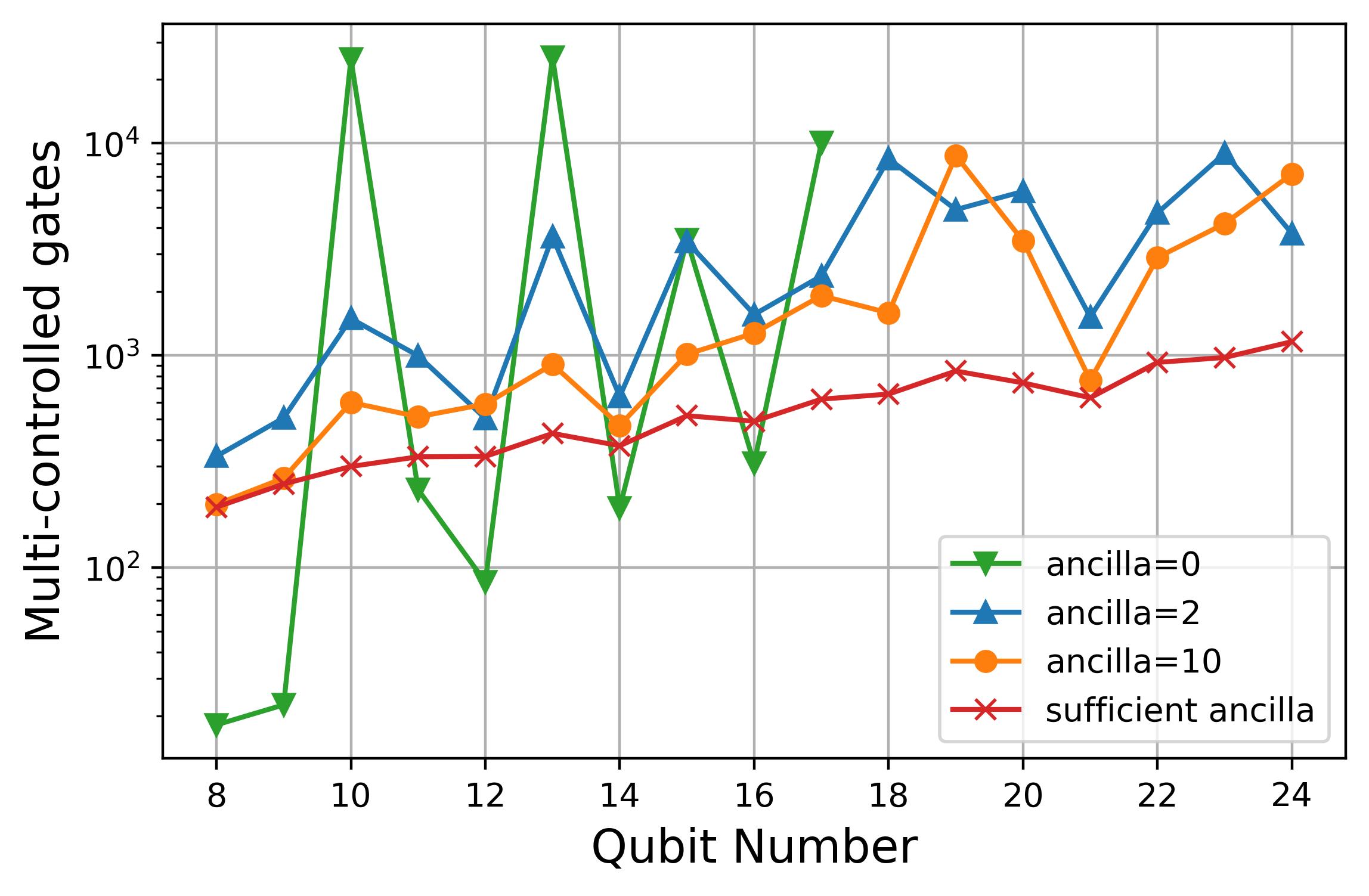}
    \includegraphics[scale=0.6]{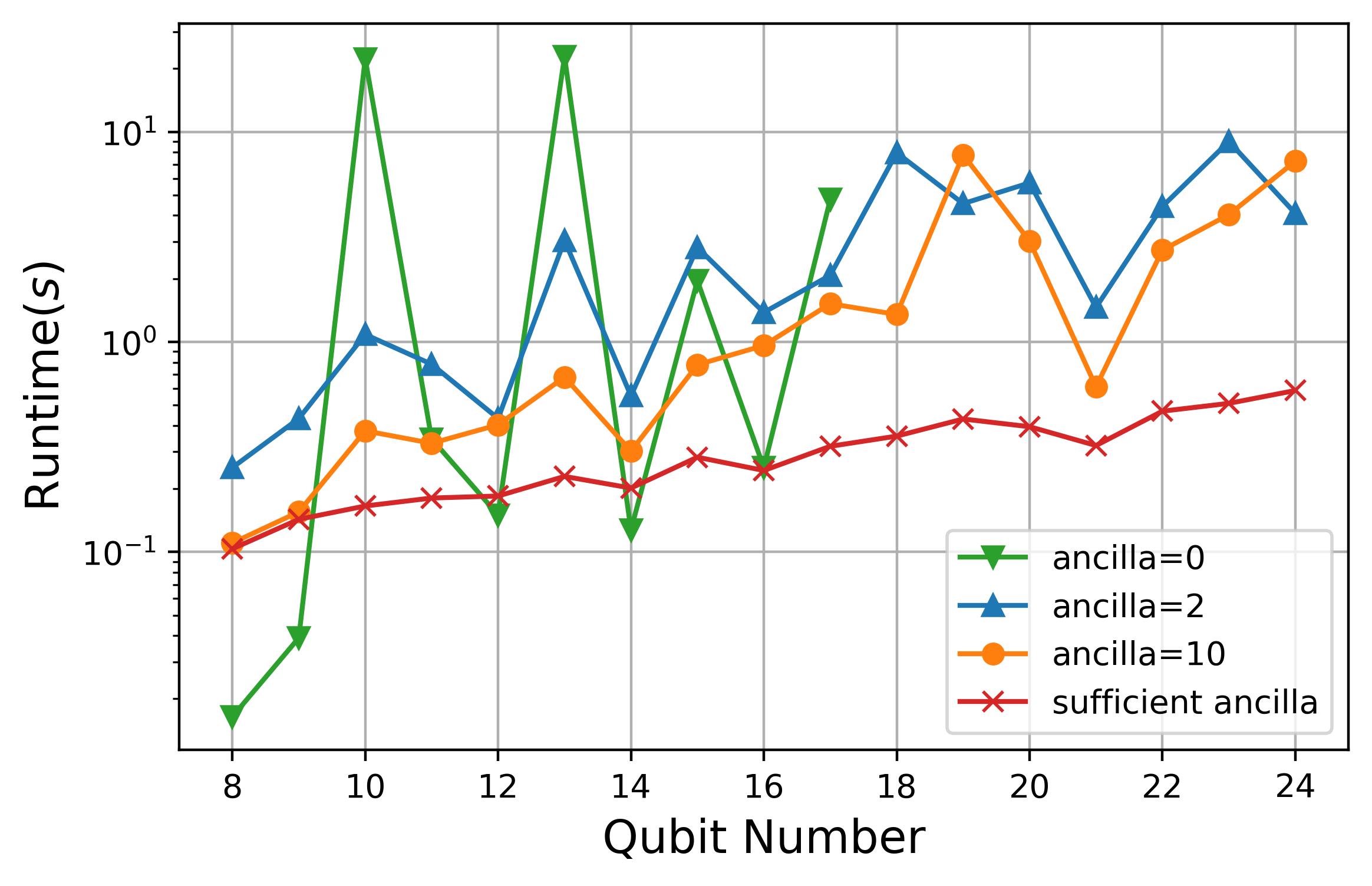}    
    \caption{Compare with other ancilla-based LimTDD algorithms.}
    \label{fig:vs_ancilla}
\end{figure*}

Finally, we compared the performance of all four algorithms proposed in our work. For Alg. \ref{alg:State_Pre_lima}, we set the number of auxiliary qubits to 10. Fig. \ref{fig:vs_ancilla} shows the runtime and multi-qubit gate complexity of the algorithms. The number of gates is the one after transpilation.

\begin{itemize}
    \item \textbf{Gate Complexity}:
    Alg. \ref{alg:State_Pre_noa} exhibits unstable performance, mainly due to the large number of multi-control bit gates introduced during compilation. Excluding Alg. \ref{alg:State_Pre_noa}, the number of multi-qubit gates required for state preparation decreases with an increased number of ancilla qubits. Furthermore, the trend of gate complexity becomes more stable and smooth as more ancilla qubits are utilised.

    \item \textbf{Runtime Complexity}: The runtime trend mirrors the gate complexity trend. Excluding Alg. \ref{alg:State_Pre_noa}, the runtime decreases with the number of ancilla qubits.

\end{itemize}

\section{Conclusion}\label{sec:conclusion}

In this paper, we proposed novel quantum state preparation algorithms based on the Local Invertible Map Tensor Decision Diagram (LimTDD). These algorithms significantly improve the efficiency and reduce the complexity of quantum circuits, particularly for large-scale quantum states. The compact representation of LimTDD enables substantial improvements in both time and gate complexity, achieving exponential efficiency gains in certain scenarios.

Our experiments demonstrate that the proposed methods outperform existing frameworks such as Qiskit and QuICT, especially as the number of qubits increases. The integration of LimTDD into quantum state preparation highlights its potential for handling complex or large-scale quantum states with fewer resources. This work not only advances the state-of-the-art in quantum state preparation but also provides a robust foundation for future developments in quantum computing technologies.

Future work will focus on further optimising LimTDD and exploring its applications in other quantum computing tasks. We aim to integrate our algorithms into widely used quantum computing frameworks to standardise and accelerate the quantum state preparation process. Additionally, we plan to develop more sophisticated interfaces to streamline the workflow for quantum physicists and experimentalists, enabling them to efficiently generate the quantum states required for their research.

In summary, this paper has demonstrated the potential of LimTDD in quantum state preparation, paving the way for more efficient and scalable quantum computing solutions.







\bibliographystyle{IEEEtran}
\bibliography{reference}

\end{document}